\documentclass[twocolumn,trackchanges]{aastex7}
\usepackage{hyperref}
\usepackage{placeins}

\received{February 5, 2026}
\revised{May 1, 2026}
\accepted{May 12, 2026}
\submitjournal{AJ}

\begin{document}

\title{Near UV Stellar Activity and Brightness Fluctuations of the Alpha Centauri AB Star System from Weeks to Decades~--~Inputs for Reflected Light Spectroscopy with HWO}

\author[orcid=0000-0003-1375-7101]{Dolon Bhattacharyya}
\affiliation{Laboratory for Atmospheric and Space Physics, University of Colorado Boulder, Boulder, CO 80303}
\email[show]{dolon.bhattacharyya@lasp.colorado.edu}  

\author[orcid=0000-0002-1002-3674]{Kevin France} 
\affiliation{Laboratory for Atmospheric and Space Physics, University of Colorado Boulder, Boulder, CO 80303}
\affiliation{Center for Astrophysics and Space Astronomy, University of Colorado Boulder, Boulder, CO 80303}\email{kevin.france@colorado.edu}

\author{Soumit Rao}
\affiliation{Department of Physics, Ashoka University, Sonipat, Haryana 131029, India}\email{soumit.rao_ug2023@ashoka.edu.in}

\author{Sebastian Escobar} 
\affiliation{Laboratory for Atmospheric and Space Physics, University of Colorado Boulder, Boulder, CO 80303}\email{sebastian.escobar@colorado.edu}

\author[orcid=0000-0001-9667-9449]{David J. Wilson} 
\affiliation{Laboratory for Atmospheric and Space Physics, University of Colorado Boulder, Boulder, CO 80303}\email{david.wilson@colorado.edu}

\author[orcid=0000-0002-4701-8916]{Arika Egan} 
\affiliation{Applied Physics Laboratory, Johns Hopkins University, Baltimore, MD 21218}\email{arika.egan@jhuapl.edu}

\author[orcid=0000-0003-4372-7405]{Phillip Chamberlin} 
\affiliation{Laboratory for Atmospheric and Space Physics, University of Colorado Boulder, Boulder, CO 80303}\email{phil.chamberlin@lasp.colorado.edu}

\author[orcid=0000-0002-4166-4263]{A. G. Sreejith} 
\affiliation{Space Research Institute, Austrian Academy of Sciences, Schmiedlstrasse 6, 8042 Graz, Austria}\email{Sreejith.Aickara@oeaw.ac.at}

\author[orcid=0000-0003-2631-3905]{Alexander Brown} 
\affiliation{Center for Astrophysics and Space Astronomy, University of Colorado Boulder, Boulder, CO 80303}\email{alexander.brown@colorado.edu}

\begin{abstract}

We present the most comprehensive near-ultraviolet ({NUV: 2550-3255 \r{A}}) activity record to date for the Alpha Centauri AB system, combining archival IUE and HST observations spanning nearly five decades with new high-cadence CUTE measurements. We show that Alpha Centauri A exhibits predominantly quiescent NUV behavior, with the majority of observations remaining within $\pm1\sigma$ of the median flux and only rare chromospheric flaring events ($\sim$1 flare every 12 years), consistent with its weak chromospheric activity and 19-year stellar cycle inferred from X-ray and FUV observations. In contrast, Alpha Centauri B displays a broader variability envelope, characterized by more frequent and higher-amplitude chromospheric excursions that track its well-established $\sim$8-yr magnetic activity cycle. Using Lomb–Scargle analysis on the Mg II index derived from CUTE observations, we estimate the rotational period of Alpha Centauri A to be on timescales of $\sim$15–20 days. We also confirm the coherence of the stellar activity cycle of Alpha Centauri B in the NUV with its X-ray activity cycle. These data establish a critical reference framework for interpreting reflected-light observations of terrestrial exoplanets and for assessing the detectability of ozone and other biosignature-related features at NUV wavelengths with future facilities such as the Habitable Worlds Observatory. These results indicate that HWO observations of terrestrial exoplanets in reflected light photometry and spectroscopy around magnetically inactive early G-type stars and early K-type stars may be expected to show 10~--~20\% and 30~--~40\% temporal flux variability, respectively, over the course of months to years from the changing stellar inputs alone.  
\end{abstract}



\section{Introduction} 


Exoplanetary science has witnessed an explosion of discovery in the last three decades, with over 6000 exoplanets with basic properties (e.g., mass and radius) now known \citep{bryson20}. The next task is to characterize the composition and evolution of their atmospheres, and ultimately determine which of these worlds support habitable conditions. Discovery in this area is not only at the forefront of modern astrophysics and planetary science, it has broad-reaching public interest in our quest to {\it answer the fundamental question `Are we alone?'}.  The National Academies of Sciences, Engineering, and Medicine 2020 Decadal Survey on Astronomy and Astrophysics ``Pathways to Discovery'' (Astro2020; \citealt{2020Astro}) report prioritized a large IR/optical/UV space telescope to pursue ambitious science goals under the Survey's ``Worlds and Suns in Context'', ``Cosmic Ecosystems'', and "New Messengers and New Physics" themes. This mission, now adopted by NASA as the ``Habitable Worlds Observatory'' (HWO;\citealt{nasa_hwo}), combines these three themes into a facility that will enable a broad and transformative science portfolio.  One of the signature science goals laid out by \citeauthor{2020Astro} (\citeyear{2020Astro}) was the detection and characterization of at least 25 rocky planets in the liquid water habitable zones around nearby stars.  This charge can be split into two questions: (1) what is the frequency of terrestrial planets that display signatures of biological processes, and (2) how does this frequency evolve with the properties of their star-planet systems? 


Our ability to answer the first question depends directly on the performance of the high-contrast imaging system on HWO \citep{2026Feinberg}.  The minimization of the inner working angle and ability to achieve contrast levels $<$ 10$^{-10}$ enables the largest number of habitable zones to be searched, spanning the widest range of stellar spectral type.  The second question is connected to the UV capabilities and  the compatibility of the optical coatings with the coronagraph, which determines our ability to explore atmospheric conditions as a function of both time and host star mass.   

\subsection{Habitable Worlds in Time: NUV Reflected Light Observations of Atmospheric Oxygen Content  }\label{section-NUV}
As Earth's atmospheric composition has evolved over its 4.5 billion year history, so too have the spectral signatures of the primary atmospheric constituents and signs of active biology.  The abundance of oxygen in Earth's atmosphere is a prominent manifestation of our biosphere and an important indicator of the planet's overall redox state, which is an essential chemical context for interpreting biosignatures.  Although modern Earth has strong O$_2$ absorption features in the visible, this was not the case during $\approx$80\% of Earth's 4.5-billion-year history.

During the Hadean and Archaen period of Earth's history, $>$ 2.5 billion years ago (Ga), oxygen was a minor species and higher abundances of CO$_{2}$ and CH$_{4}$ (probed by spectral features in the 0.7~--~1.8 $\mu$m range) reveal the major atmospheric composition, including the possibility of an anaerobic surface biology (see, e.g, Figure 1.1 in Astro2020 \citep{2020Astro};~\citealt{Krissansen-Totton18,arney16}).  Photosynthetic life dates back to $\approx$3.4 Ga to 2.4 Ga, with the onset of the Great Oxidation Event (GOE).  However, atmospheric O$_2$ levels remained under 10$\%$ present atmospheric level (PAL)  until $\approx$0.8 -- 0.5 Ga (the Proterozoic eon, $\approx$0.5 -- 2.5 Ga, brackets the period between the GOE and the later rise of atmospheric O$_2$ to PAL). These levels of atmospheric O$_2$ are not expected to be detectable by HWO in the visible (see Figure 1.1 in Astro2020 \citep{2020Astro}).  Instead, near-UV absorption of O$_3$, a highly specific photochemical product of O$_2$, can be an indirect indicator of molecular oxygen.  Ozone's Hartley-Huggins band (210 -- 380~nm) in the near-UV shows deep absorption in Proterozoic-Earth models, even with 10$^{-3}$ PAL O$_2$.  A recent study \citep{damiano2023} indicated that low-resolution spectroscopy with a short-wavelength cutoff of 250~nm enables the recovery of O$_3$ abundance under Proterozoic scenarios. Taken together, performance across the full 0.25~--~1.8~$\mu$m region captures the evolution of Earth-like atmospheres and is a major driver for  Astro2020's Worlds and Suns in Context theme (see also, \citealt{totton25}).




The O$_3$ spectra are obtained as a reflected light measurement of the orbiting planets, and therefore an understanding of the near-UV stellar flux is essential to predicting the detectability of the O$_3$ signal and its eventual retrieval in reflected light spectroscopy. Limitations in our ability to predict the NUV spectrum of the planet hosting star, and how that spectrum might vary during an exoplanet observation, will translate into a direct limitation on our ability to predict and interpret those reflected light spectra. Variations in the illuminating stellar spectral energy distribution could be a significant source of uncertainty in quantifying the reflected light signal from orbiting planets, however, very little is known about the spectroscopic variation of cool stars in the 250~--~350 nm NUV band where the photospheric and chromospheric contributions to the stellar light both begin to contribute for many cool stars~\citep{loyd16}.  Stellar spectroscopic inputs have been identified as a key gap in our preparation for a UV coronagraphic mode on HWO~\citep{peacock25} and retrieved exoplanet atmospheric properties rely on a detailed understanding of the host star properties (e.g., \citealt{tuchow25}).  

In this paper, we combine new and archival observations of our nearest Sun-like neighbors, the $\alpha$ Centauri AB system in the NUV wavelength. In this work, we define the NUV wavelength range as 2550–3255 \r{A}, corresponding to the spectral coverage of the International Ultraviolet Explorer (IUE), Hubble Space Telescope (HST), and Colorado Ultraviolet Transit Experiment (CUTE) observations used here.  We take advantage of the proximity and interest in this stellar system to develop a long-term NUV stellar brightness archive of our nearest Sun-like neighbors, quantifying brightness variations on the timescales from weeks (the stellar rotation timescale) to decades and tracking changes in the chromospheric activity of the system.   These data can be incorporated into the technical analysis and data simulations of high-contrast imaging systems for HWO.  We present this new analysis and associated machine-readable data \citep{Bhattacharyya2026AlphaCenNUV} to provide empirical bounds on the stellar variations that could be expected across the multi-year observing baselines which will be carried out with HWO and how these variations are propagated to the uncertainties on the retrieved oxygen abundances of these worlds.    


\section{NUV Observations of Alpha Centauri}
The nearby Alpha Centauri star system contains three stars, Alpha Centauri A which is a G2 V star, Alpha Centauri B which is a K1 V star, and Proxima Centauri which is a M5.5 Ve star. The A and B stars orbit each other closely at a distance of $\sim35$ AU whereas Proxima Centauri is located at a distance of $\sim13,000$ AU. In this paper we have focused on studying the NUV activity of the AB star system. 

\begin{figure*}[t]
    \centering
    \includegraphics[width = \textwidth, height=9cm]{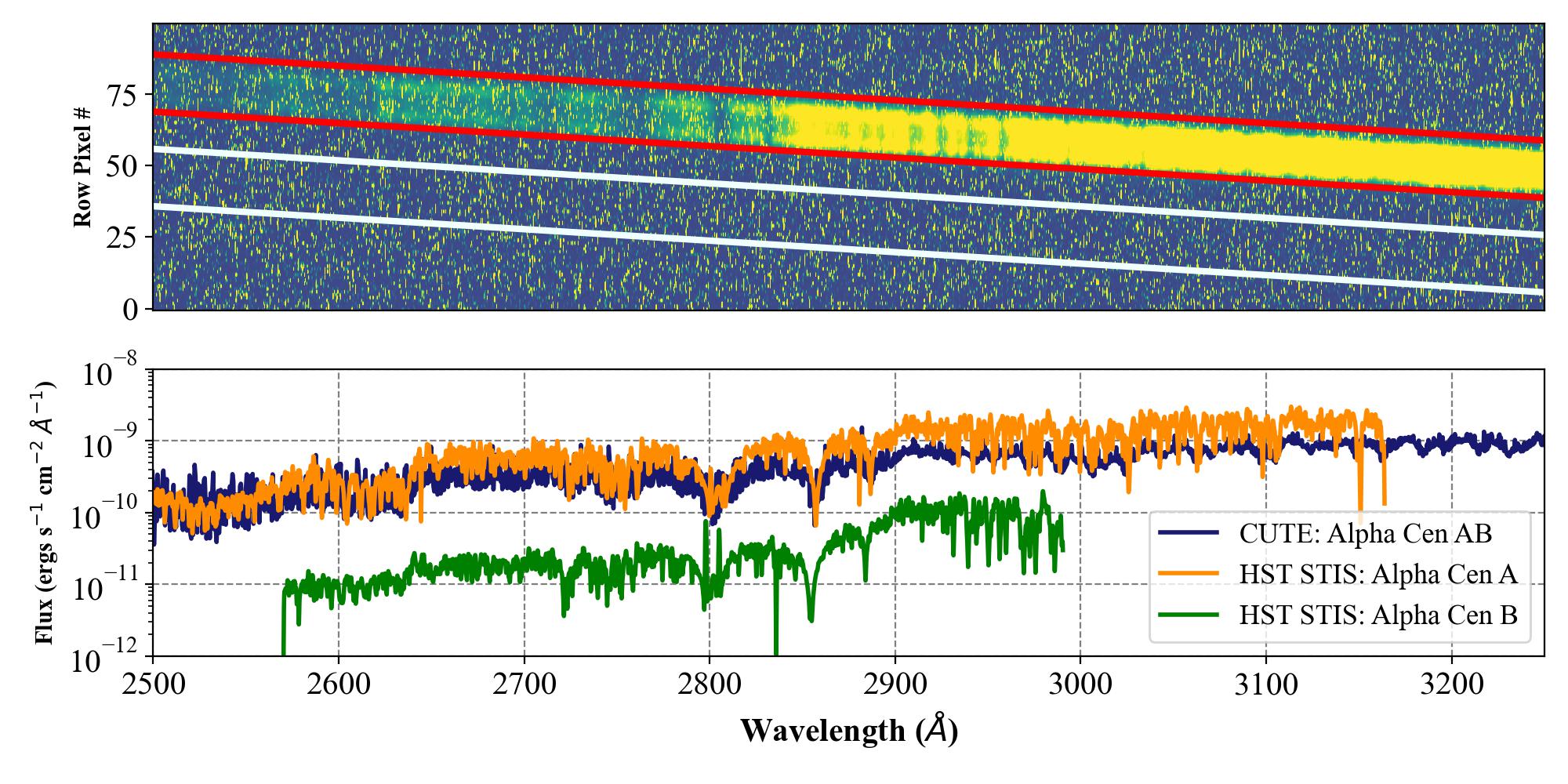}
    \caption{\textbf{Top:} CUTE NUV 2D CCD spectra of Alpha Centauri from a single 240 s exposure. The spectral extraction region is marked by the red parallel lines and the background extraction region is marked by the white parallel lines. \textbf{Bottom:} 1D background corrected spectra obtained by summing the counts from the columns in the spectral extraction region and converted into stellar flux units by multiplying with the CUTE's wavelength-dependent sensitivity curve. HST spectra of Alpha Centauri A and B are also plotted as a reference for CUTE's observed flux. The resolution of the HST spectra has been reduced to match CUTE’s resolution via convolution with a Gaussian kernel, enabling a direct comparison of the absorption features with the CUTE data.} 
    \label{fig:fig1}
\end{figure*}

The Alpha Centauri AB star system has been observed with space-based observatories like the IUE and the HST in the NUV wavelength range. These observations started as early as 1978 and have continued into 2025. The HST and IUE observations together cover a long baseline of the stellar activity displayed by the Alpha Centauri A and B stars. More recently, this star system was observed by the CUTE \citep{2023France} CubeSat in the NUV. The CUTE observations were executed over a 1.5 month period in 2024 and a 2.5 month period in 2025 and have the highest NUV time cadence since the turn of the century. Unlike HST and IUE, CUTE cannot spatially resolve the two stars. However, Alpha Centauri A is approximately an order of magnitude brighter than Alpha Centauri B at NUV wavelengths when B is near activity minimum (Fig. \ref{fig:fig1}), consistent with its state during the 2024–2025 CUTE observations (\citealt{Ayres_2023}). Hence, about $90\%$ of the flux measured by CUTE originates from Alpha Centauri A, allowing the observations to be interpreted as a proxy for the NUV variability of Alpha Centauri A. In this paper, we present the analysis of the IUE, HST, and CUTE data sets to characterize the NUV activity of the Alpha Centauri AB system with a long temporal baseline. 

\subsection{CUTE Observations of Alpha Centauri}
The CUTE mission contains a spectrograph housed in a 6U CubeSat capable of obtaining NUV spectroscopy of astronomical objects \citep{2023France,2023Egan}. Its operating range is between 2480 - 3306 \r{A} with a resolving power R $\sim 750$  ($\Delta \lambda \approx 3.9 - 4.7$\r{A}) across the bandpass. The CUTE data are recorded on a passively cooled backside-illuminated CCD with 515 spatial $\times$ 2048 active spectral pixels \citep{2021Nico}. CUTE is the first and longest-running ultraviolet astrophysics CubeSat since its launch into low-Earth orbit in September 2021. 

CUTE first observed the Alpha Centauri AB system from 27 April 2024 - 4 June 2024 over a series of 4 separate visits. Each visit consisted of a sequence of bias, dark and science exposures conducted over a ~1-2 day period. The bias and dark exposures are taken both before and after the science exposures during each visit. The exposure time for individual dark and science frames is 240 seconds for all visits. The second set of observations of the Alpha Centauri system with CUTE was conducted exactly a year later from 27 April 2025 - 29 June 2025 over a series of 10 separate visits at a weekly time cadence. The exposure sequence per visit were similar to the 2024 observations except that this time it only consisted of bias and science exposures. Dark exposures were deemed unnecessary for the 2025 campaign. This is because analysis of the 2024 dataset demonstrated that background counts obtained from detector regions devoid of stellar spectra within the science exposures yield a more accurate estimate than separate dark exposures, owing to the closer correspondence in orbital conditions (Fig. \ref{fig:fig1}; \citealt{2022Sreejith}). The same technique has been applied towards the analysis of other CUTE science targets like WASP189b, and KELT-9b \citep{2023Sreejith,2024Arika}. The exposure time was kept at 240 seconds for the science frames.    

The two-dimensional NUV spectra of Alpha Centauri A recorded by CUTE in each science frame are reduced to one dimension by summing over all the pixels in the spatial direction for a reduced portion of the detector which contains the stellar spectra. The background detector counts (dark + bias) are estimated from a region of similar spatial width as the stellar spectra in the same science frame, which is not exposed to star light. Figure \ref{fig:fig1} (top) shows an example of a CUTE Alpha Centauri science exposure with the stellar extraction region marked by red parallel lines and the background extraction region marked by white parallel lines. The bottom plot contains the one-dimensional background-corrected spectra of the star obtained by summing over the spatial direction within the stellar extraction region and converted to stellar flux units by multiplying with the CUTE wavelength-dependent sensitivity curve \citep{cute2025db}. The plot also includes HST spectra of Alpha Centauri A and B for comparison with the CUTE data. The HST observations of A and B were obtained on 10 April 1999 and 24 January 2016, respectively. These datasets were selected for Fig.~\ref{fig:fig1} because they provide the broadest wavelength coverage within the CUTE bandpass, and the activity state of B is comparable to the 2024–2025 CUTE observations, corresponding to a near-minimum phase \citep{2023Ayres}. Alpha Centauri A exhibits only weak variability in the FUV and NUV, and is therefore largely insensitive to the choice of observation date (\citealt{2023Ayres}; this work).

\subsection{HST Observations of Alpha Centauri}
HST observed Alpha Centauri A and B individually with the Space Telescope Imaging Spectrograph (STIS) instrument in the NUV wavelength range with the high-resolution E230H grating with a resolving power of $\sim$114,000 and a wavelength coverage of 1626 - 3159 \r{A}. Different combinations of apertures and neutral density filters were used to make the observations. Alpha Centauri A was observed intermittently from 1999 to 2024 while Alpha Centauri B was observed intermittently from 2010 to 2024. These observations were from different general observer programs with science objectives which include studying the magnetic and coronal activity cycles of the A and B stars by observing different emission lines \citep[and references therein]{2023Ayres}. As a result, the central wavelength of the grating was varied to cover different portions of the 1626 - 3159 \r{A} wavelength band at different times. Due to this, we do not have consistent spectra of the two stars in the CUTE bandpass from the observation years 1999 - 2024. However, we do have spectral coverage over the Mg II h and k chromospheric emission lines at 2803.53 \r{A} and 2796.35 \r{A} for the entire time period of the HST observations. Hence we are able to map out the temporal variation of the chromospheric activity of the two stars from the HST dataset. Figure \ref{fig:fig1} displays a sample HST STIS spectrum from Alpha Centauri A and B that has been convolved with a Gaussian kernel to degrade its native high resolution to that of CUTE, enabling a direct comparison of spectral features across the two instruments. The wavelength coverage does not exceed 3159 \r{A} on the redder end of the spectrum for Alpha Centauri A, which is less than the CUTE's longest wavelength limit of 3306 \r{A}. At the same time, Alpha Centauri B's wavelength coverage stops short of 3000 \r{A}. The plotted data for A and B has the largest coverage in wavelength space within the CUTE bandpass for both stars among the entire HST data set used in this analysis (\href{https://doi.org/10.17909/dk8s-n214}{doi:10.17909/dk8s-n214}).  

\subsection{IUE Observations of Alpha Centauri}
IUE observed Alpha Centauri A and B with the Long Wavelength Prime/Long Wavelength Redundant (LWP/LWR) cameras with a high-resolution (R $\sim$10,000 - 15,000 across the NUV bandpass) UV spectrograph paired with either a large aperture slit of 10$^{\prime\prime} \times 20^{\prime\prime}$ in size or a small circular aperture of diameter 3$^{\prime\prime}$. IUE also used the low resolution UV spectrographic mode (6.0 \r{A} - 7.0 \r{A}) together with either of the two apertures (large/small) to observe the two stars.The wavelength coverage of the low resolution mode was 1900 \r{A} - 3200 \r{A} and the spectral resolution ranged from 0.1 \r{A} - 0.3 \r{A} within the bandpass. The IUE observations span from 1978 to 1995. 

The IUE data hosted in the Mikulski Archive for Space Telescopes (MAST) were considered for this study (\href{https://doi.org/10.17909/p38z-4y59}{doi:10.17909/p38z-4y59}). The dataset contains a total of 129 observations in the NUV bandpass of 1900 \r{A} - 3200 \r{A}. Out of the 129 observations, 68 are of Alpha Centauri A and 61 are of Alpha Centauri B. These observations were executed in the high dispersion + large aperture mode or low dispersion + large/small aperture mode. For this analysis we have looked at 47 individual spectra of Alpha Centauri A and 52 individual spectra of Alpha Centauri B obtained in the high dispersion + large aperture configuration.  21 IUE observations of Alpha Centauri A were not used in the current analysis.  Of the 21 observations, 6 are in the low resolution mode (5 with the large aperture and 1 with the small aperture). For the low resolution data, 3 have spectra with registered flux for Alpha Centauri A in the nominal range. The remaining 3 contain flux much lower than is usually recorded for the star. The other 15 of the 21 observations, obtained in high-resolution + large-aperture mode, were not used in the analysis because they exhibited either lower-than-normal recorded flux levels or had missing data in portions of the spectral band. 8 of the 15 observations were conducted between May - July 1995, which was near the end of the mission when IUE attitude control was becoming intermittent. The pointing issues would likely have degraded the star's ideal alignment within the slit for these observations. We decided to not use the three good low resolution observations (all obtained on 12 January 1980) in our current analysis to preserve the consistency of the dataset, which currently uses the most favored mode, high dispersion in conjunction with the large aperture. The high dispersion + large aperture dataset also has well-defined defined spectral windows necessary for the calculation of the Mg II index, unlike the low dispersion mode (see Section \ref{sec:mgii}).

In the current analysis, 9 observations of Alpha Centauri B present in the IUE archive were ignored. All spectra were obtained in the high resolution + large aperture mode, except for one, which was in low resolution + large aperture mode. The reason for excluding these 9 datasets are the same as Alpha Centauri A, i.e., lower-than-normal recorded flux or missing data for certain wavelength ranges. However, unlike Alpha Centauri A, the 9 observations are distributed over the entire time span of IUE's lifetime (1978 - 1995). The complete list of IUE files for Alpha Centauri A and B that have been used/not used in the analysis is given in \cite{Bhattacharyya2026AlphaCenNUV}.

In summary, we have used a total of 75 observations, each corresponding to a unique observation date, from IUE, HST, and CUTE to establish a timeline of the chromospheric activity for Alpha Centauri A by looking at the flux changes around the 2800 \r{A} MgII emission feature. The same has been done for Alpha Centauri B but without CUTE with a total of 63 unique observations. We have used the IUE and the CUTE data to analyze the temporal changes experienced by the photosphere in the wavelengths shortward and longward of the Mg II h and k lines. The same has been executed for Alpha Centauri B but without CUTE.

\section{NUV Activity of the Alpha Centauri AB Star System}
We determine the NUV stellar activity of the Alpha Centauri A and B stars over a long temporal baseline (1978 to 2025) by splitting the stellar spectra from CUTE, HST, and IUE into three wavelength bands: (1) wavelengths $< 2794$ \r{A} representative of the photospheric activity and hereby referred to as photosphere 1, (2) wavelengths 2794 \r{A} - 2805 \r{A} that capture the Mg II h and k emission lines, a tracer of chromospheric activity in stars, hereby designated as chromosphere, and (3) wavelengths $> 2805$ \r{A} representative of the photospheric activity in stars, hereby referred to as photosphere 2 in the paper. The wavelength bands used for each dataset are summarized in Table \ref{tab:bands}. The flux in these wavelength bands are then integrated to obtain the total flux received from the star corresponding to each band from the individual exposures. Examining these fluxes as a function of time reveals the temporal variability in the NUV activity of the Alpha Centauri AB stars.

\begin{figure}
    \centering
    \includegraphics[width = \linewidth]{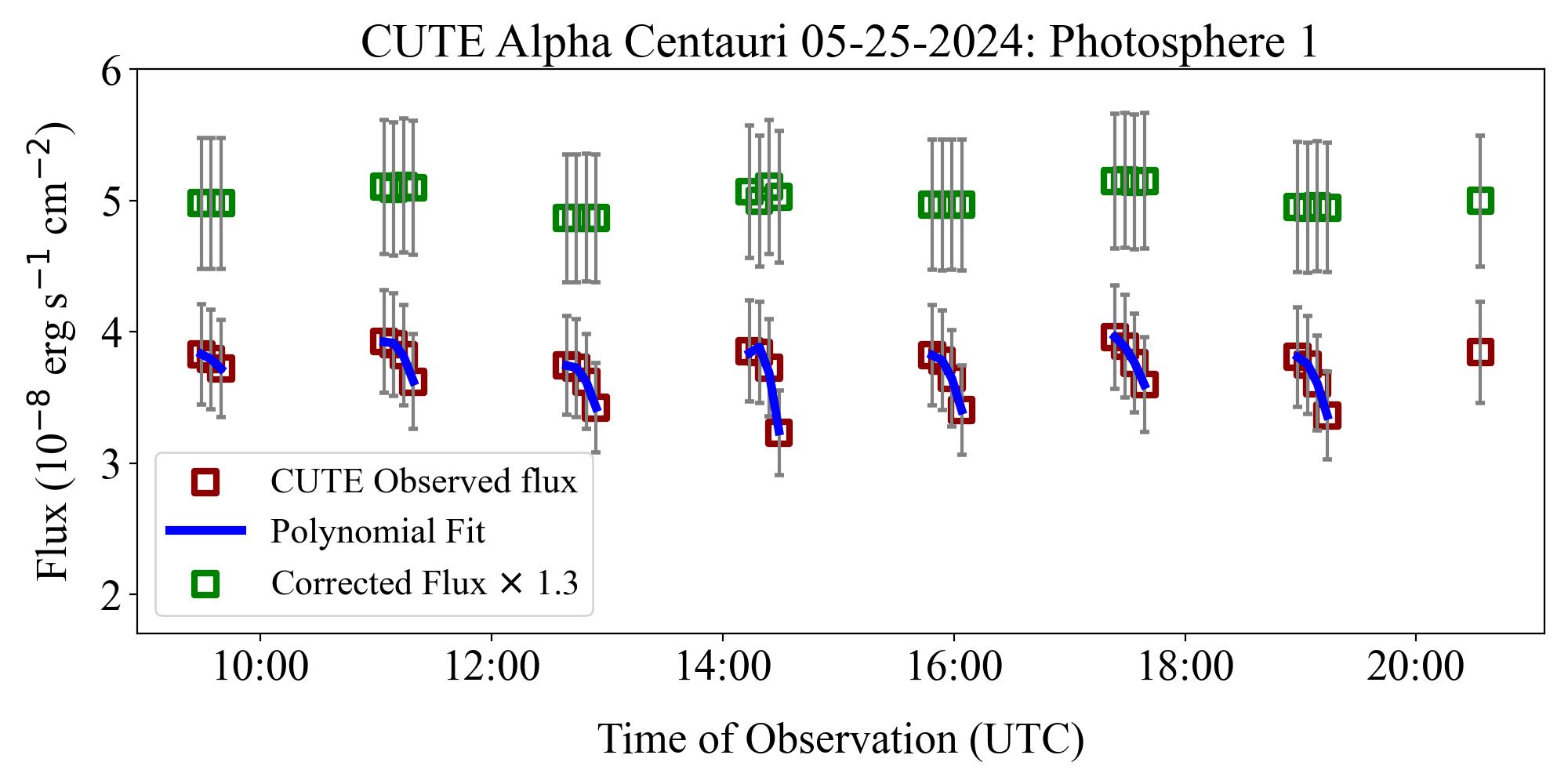}
    \caption{This figure displays the orbital correction applied to CUTE's recorded photosphere 1 integrated flux for Alpha Centauri A from the observation conducted on 25th May 2024.} 
    \label{fig:fig2}
\end{figure}

Within a single CUTE orbit, the integrated flux in the three wavelength bands representing photosphere 1, chromosphere, and photosphere 2 exhibits systematic variations driven by orbital geometry and temperature changes in the detector electronics. We correct for these variations in the stellar flux within a single orbit by fitting a polynomial function to the observed trend and dividing it out from the orbital trend. Figure \ref{fig:fig2} presents an example of the correction applied to the photosphere 1 flux of Alpha Centauri A recorded by CUTE on 25 May 2024. Similar corrections were applied to the calculated flux of the chromosphere and photosphere 2 and for all individual visits. The error bars, estimated through standard error propagation techniques, are mainly the contribution of the uncertainty in the effective area of CUTE \citep{cute2025db}.

\begin{figure*}
    \centering
    \includegraphics{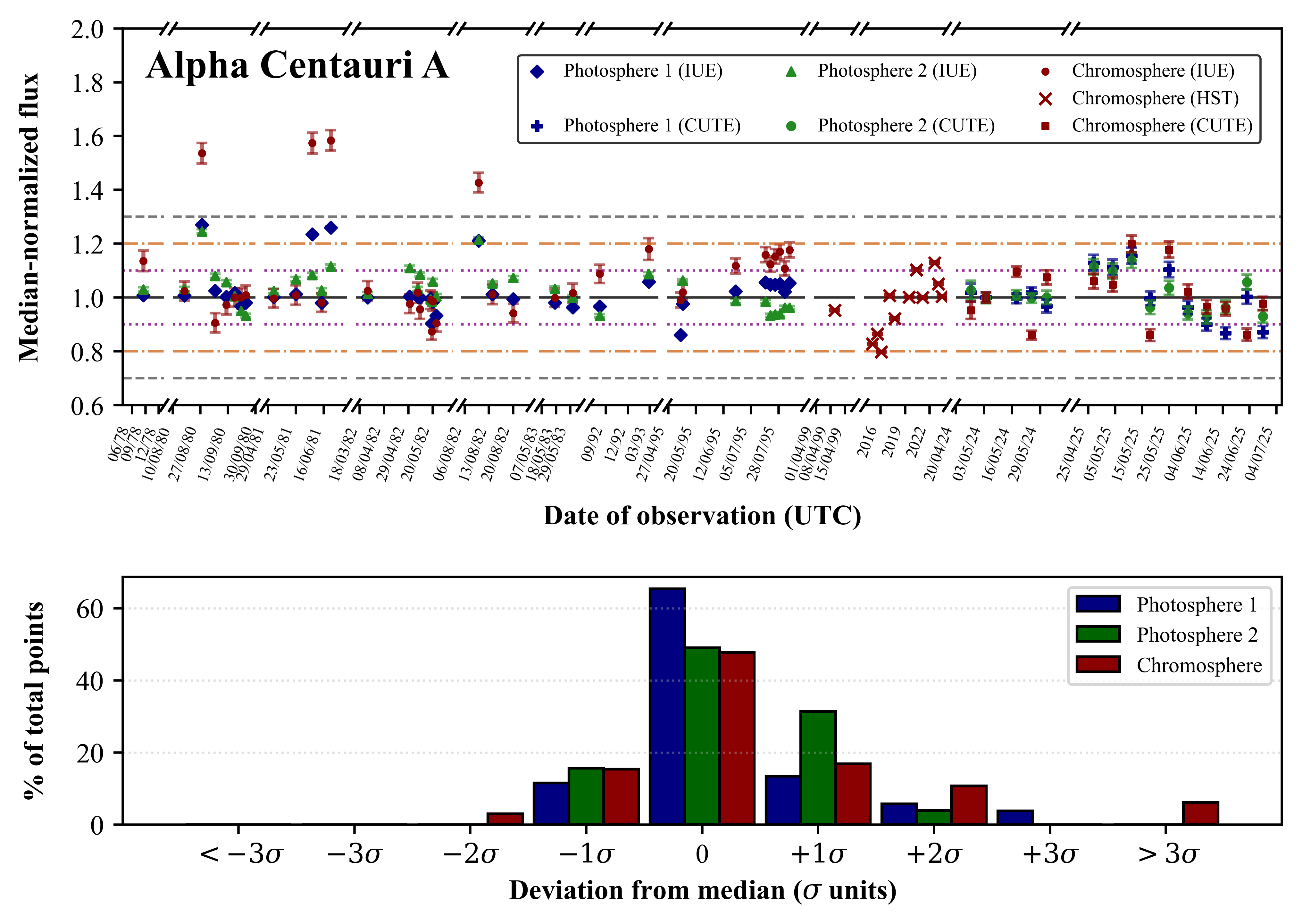}
    \caption{\textbf{Top:} This figure presents the photosphere 1, chromosphere, and photosphere 2 intensity variation of Alpha Centauri A. The x-axis represents the date of observation in the dd-mm-yy or mm-yy or yyyy format. The dotted purple line represents the $\pm1\sigma$ cutoff, the orange dot-dashed line the $\pm2\sigma$ point, and the gray dashed line the $\pm3\sigma$ cutoff point. The red points above the $+3\sigma$ line are chromospheric flaring events detected at Alpha Centauri A. \textbf{Bottom:} This figure represents the distribution of the photosphere 1, chromosphere, and photosphere 2 points around the median from CUTE, HST, and IUE. The flaring events are captured by the far right bar  above the 3$\sigma$ level.}
    \label{fig:fig_3}
\end{figure*}

\begin{table}[t]
\centering
\scriptsize
\setlength{\tabcolsep}{6pt}
\caption{Near-UV wavelength bands used in this work.}
\label{tab:bands}
\begin{tabular}{l p{3.5cm} p{2.cm}}
\hline
Component & Wavelength Range (\AA) & Physical Origin \\
\hline
Photosphere 1 & 2550--2795 & Continuum \\
Chromosphere  & 2795.5--2797.2 + 2802.68--2804.38 (IUE/HST); 2795.0--2805.0 (CUTE) & Mg II h \& k \\
Photosphere 2 & 2805--2900 (IUE/HST); 2805--3255 (CUTE) & Continuum \\
\hline
\end{tabular}
\end{table}

\begin{table}[t]
\centering
\scriptsize
\setlength{\tabcolsep}{5pt}
\caption{Median integrated near-UV fluxes used for normalization in Figure \ref{fig:fig_3} (Alpha Centauri A).}
\label{tab:medfluxA}
\begin{tabular}{l p{1.9cm} p{1.9cm} p{1.9cm}}
\hline
Instrument & Photosphere 1 & Chromosphere & Photosphere 2 \\
 & (erg s$^{-1}$ cm$^{-2}$) & (erg s$^{-1}$ cm$^{-2}$) & (erg s$^{-1}$ cm$^{-2}$) \\
\hline
IUE        & $5.80\times10^{-8}$ & $3.67\times10^{-10}$ & $3.43\times10^{-8}$ \\
HST        & --                 & $4.51\times10^{-10}$ & -- \\
CUTE 2024  & $3.88\times10^{-8}$ & $1.06\times10^{-9}$  & $1.78\times10^{-7}$ \\
CUTE 2025  & $4.95\times10^{-8}$ & $1.74\times10^{-9}$  & $1.92\times10^{-7}$ \\
\hline
\end{tabular}

\tablecomments{
{1.} HST does not provide consistent spectral coverage in the photospheric bands due to changing grating settings and neutral density filters.
{2.} The CUTE chromospheric band is wider than the IUE and HST band due to its lower spectral resolution.
}
\end{table}

The MAST archive contain observations of Alpha Centauri A and B with HST-STIS in the NUV wavelength range. For Alpha Centauri A observations, consistent coverage of the wavelengths representative of photosphere 1 and photosphere 2 were not present in the archive. For Alpha Centauri B the longer wavelengths covered by the photosphere 2 portion of the spectrum was sometimes paired with a neutral density filter. This resulted in inconsistent intensities recorded by HST for $\lambda>2850$ \r{A} throughout the dataset. The only consistency in coverage in the MAST archive existed around the Mg II h and k lines for Alpha Centauri A and B. Hence we have used the HST dataset to trace the temporal variability in chromospheric activity of the two stars. Uncertainty in the integrated flux over the 2794 \r{A} - 2805 \r{A} band is estimated through standard error propagation techniques from the recorded error in flux present in the MAST archive. Additionally, a 10$\%$ error representative of the uncertainty in absolute calibration of HST-STIS E230H and target centering within the slit is also included in the final error value \citep{STIS_IHB}. HST has the longest temporal coverage ($\sim 25$ years) for Alpha Centauri A. 

The IUE data in the MAST archive contain separate records for Alpha Centauri A and B observations. The temporal coverage for both stars is $\sim17$ years from 1978 - 1995. The spectra were consistent over the photosphere 1, chromosphere, and photosphere 2 bandpass. Hence, the IUE dataset was ideal for studying the variation in NUV activity of the two stars over a long temporal baseline. Error in the integrated flux for photosphere 1, chromosphere, and photosphere 2 for the IUE data is estimated via standard error propagation techniques from the error in flux present in the MAST archive. An additional $10\%$ uncertainty is included in the final estimated error value in the integrated flux to represent the uncertainty in absolute calibration of IUE's NUV channel. 

\begin{figure}
    \centering
    \includegraphics[width = \linewidth]{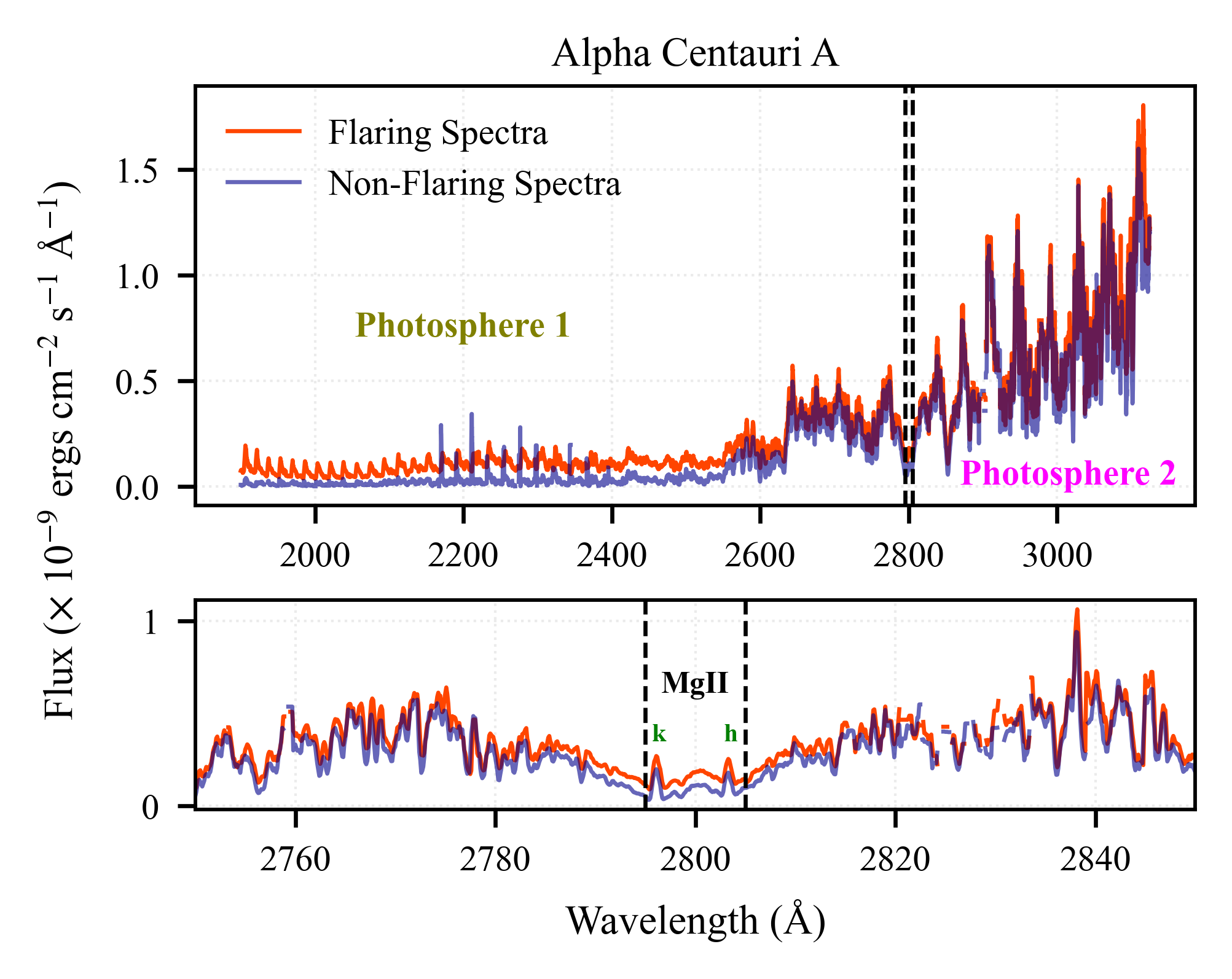}
    \caption{\textbf{Top:} This figure compares a flaring spectrum vs a normal spectrum of Alpha Centauri A as observed by IUE. The difference in flux is more noticeable in the shorter wavelength regime labeled as Photosphere 1. \textbf{Bottom:} This figure zooms in on the region marked by the dashed lines in the top plot containing the Mg II h and k lines which are proxies for Chromospheric activity in a star. During a flare the intensity of the h and k lines also increase.} 
    \label{fig:fig4}
\end{figure}

\begin{table}[t]
\centering
\scriptsize
\setlength{\tabcolsep}{5pt}
\caption{Median integrated near-UV fluxes used for normalization in Figure \ref{fig:fig_5} (Alpha Centauri B).}
\label{tab:medfluxB}
\begin{tabular}{l p{1.9cm} p{1.9cm} p{1.9cm}}
\hline
Instrument & Photosphere 1 & Chromosphere & Photosphere 2 \\
 & (erg s$^{-1}$ cm$^{-2}$) & (erg s$^{-1}$ cm$^{-2}$) & (erg s$^{-1}$ cm$^{-2}$) \\
\hline
IUE & $5.07\times10^{-9}$ & $1.85\times10^{-10}$ & $3.80\times10^{-9}$ \\
HST & -- & $1.67\times10^{-10}$ & -- \\
\hline
\end{tabular}

\tablecomments{HST does not provide consistent spectral coverage in the photospheric bands for Alpha Centauri B due to instrumental configuration.}
\end{table}

\begin{figure*}
    \centering
    \includegraphics[width = \linewidth]{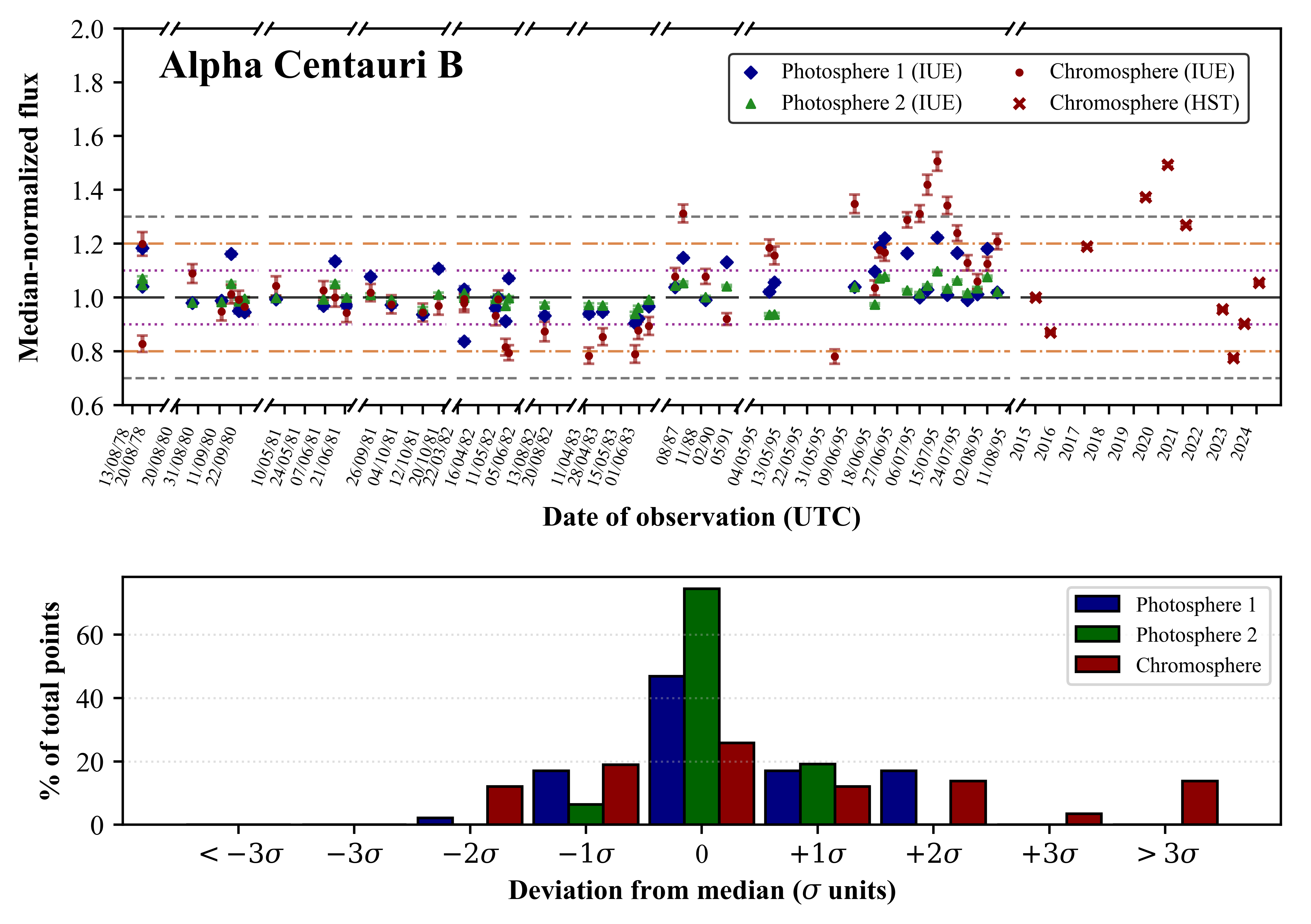}
    \caption{\textbf{Top:} This figure presents the photosphere 1, chromosphere, and photosphere 2 intensity variation of Alpha Centauri B. The x-axis represents the date of observation in the dd-mm-yy or mm-yy or yyyy format. The dotted purple line represents the $\pm1\sigma$ cutoff, the orange dot-dashed line the $\pm2\sigma$ point, and the grey dashed line the $\pm3\sigma$ cutoff point. \textbf{Bottom:} This figure represents the distribution of the photosphere 1, chromosphere, and photosphere 2 points around the median from HST and IUE.}
    \label{fig:fig_5}
\end{figure*}

Figure \ref{fig:fig_3} shows the changing NUV activity of Alpha Centauri A by tracing the variations in the photosphere 1, chromosphere, and photosphere 2 activity with time over a period of 47 years starting from 1978 to the present. The top plot shows the median-normalized integrated flux received from the star with time, while the bottom plot shows the distribution of the wavelength-integrated photospheric and chromospheric intensity around the median. As seen from the histogram plot, the distribution of points resembles a skewed Gaussian with most points located around the median. Table \ref{tab:medfluxA} contains the median flux values corresponding to each spacecraft used to normalize the data.

The red points $>+3\sigma$ are from the large increases in intensity recorded for the Mg II h and k lines, which are proxy for the chromospheric activity of the star. The increase in chromospheric activity is accompanied by an increase in the intensity of the photosphere 1 flux ($\lambda < 2794$ \r{A}). This is expected for classical optical/UV  flares on cool stars \citep{2009K_flares,2010K_flares,2019K_flares,Kleint2016ApJ81688,Kerr2015AA582A50}. Figure \ref{fig:fig4} compares the flaring spectrum of Alpha Centauri A as observed by IUE on $6^{th}$ June 1981 vs a non-flaring spectrum as recorded by IUE on $14^{th}$ June 1981. Only 4 distinct flaring signatures are detected for Alpha Centauri A in the entire dataset presented here. This is consistent with Alpha Centauri A being a magnetically quiet G dwarf, whose near-UV emission is dominated by relatively stable photospheric flux (Fig. \ref{fig:fig_3}), and exhibits only low-amplitude chromospheric variability compared to the more active K- and M-type stars \citep{Ayres_2015,2019K_flares}. The four flares were also recorded between August 1980 and August 1982. This is close to the star's maxima in its activity cycle, which would have occurred at the end of 1978 considering an activity cycle of 19 years estimated from X-ray observations \citep{2023Ayres}.  

Alpha Centauri A has been observed in the X-ray through long-term observing campaigns with Chandra and XMM-Newton as well as in the far-ultraviolet with HST \citep{Ayres_2009, Ayres_2015,Ayres_2023}. The X-ray observations revealed a 19 year stellar cycle for the star, which is longer than the 11 year cycle for the Sun, even though both stars are of a similar spectral type. According to the X-ray data Alpha Centauri A is currently approaching its minimum in the stellar cycle with its past maxima being recoded within the 2010 - 2017 time frame \citep{Ayres_2023}. The FUV flux from Alpha Centauri A as observed by HST, on the other hand, remained steady between 2010 to 2017 with little change associated with the maxima in the X-ray cycle. Only the Fe XII coronal forbidden line at $\lambda = 1242$ \r{A} partly mirrored the slow rise in X-ray activity between 2010 - 2017 \citep{Ayres_2015}. This is in line with the changes in NUV activity presented in Figure \ref{fig:fig_3} wherein most of the photospheric and chromospheric intensity received from the star is located around the median of the distribution.

Figure \ref{fig:fig_5} shows the NUV activity of Alpha Centauri B with time from 1978 to 2024. The median flux used to normalize the different datasets is given in Table \ref{tab:medfluxB}. Alpha Centauri B is of spectral type K1V. These type of stars are much more active in the shorter wavelengths than G-type stars like Alpha Centauri A. This is evident in the more symmetric distribution of points representative of the photospheric and chromospheric activity in the star around the median with more points present in the $\pm2\sigma$ bin in Figure \ref{fig:fig_5} as compared to Figure \ref{fig:fig_3}. Overall, Alpha Centauri B's distribution of chromospheric activity is closer to symmetric about the median than that of Alpha Centauri A as quantified by the skewness values of -0.14 for B vs -0.33 for A respectively. 

Like Alpha Centauri A, B has also been observed in the X-ray and FUV wavelengths \citep{Ayres_2023,Ayres_2015}. The X-ray activity of the star indicates that the stellar cycle has a period of $\sim$8 years. The last two maxima were determined to be in 2012 and 2020. This exactly coincides with the increase in chromospheric activity of the star marked by the red circles and the red crosses in Figure \ref{fig:fig_5}. Similarly, low chromospheric activity represented by the red circles in the year 1983 coincides with the minima in the stellar cycle of Alpha Centauri B. According to \cite{Ayres_2015} the FUV activity of B mimics the X-ray activity cycle of the star, unlike Alpha Centauri A. Our findings corroborate this trend with the extension to NUV wavelengths.    

\section{Stellar Activity Cycle and Rotation Period}
\label{sec:mgii}

The Mg II h and k lines are widely used diagnostics of stellar chromospheric variability \citep{B&M}. In particular, the Mg II core to wing ratio provides a sensitive measure of the star's activity cycle. For the Sun, the Mg II index defined by \cite{HandS}, later modified for higher resolution instruments, has been widely used to monitor its chromospheric activity based on data from various missions such as SBUV, SOLSTICE, SUSIM, and GOME \citep{1999Viereck,2019ESnow,Donnelly1988_AdvSpaceRes,Skupin2005_ESASP572}. We took advantage of the presence of the Mg II lines for Alpha Centauri A and B in our NUV dataset from IUE, HST, and CUTE to derive the Mg II index in order to study the long-term chromospheric activity of the stars. 

For Alpha Centauri A and B the Mg II index was calculated based on the formula:
\begin{equation}
\label{eq:mgii_index}
\text{Mg\,II index} =
\frac{I_{k} + I_{h}}{I_{\text{blue-wing}} + I_{\text{red-wing}}}
\end{equation}

In equation \ref{eq:mgii_index} the term $I_{k}$ and $I_{h}$ represents integrated flux within two 1.70 \r{A} wide windows centered around 2803.53 \r{A} and 2796.35 \r{A}. Similarly, the terms $I_{\text{blue-wing}}$ and $I_{\text{red-wing}}$ in the equation refer to two 15 \r{A} wide windows centered at 2770.50 \r{A} and 2817.50 \r{A} respectively. The choice of these windows works well for the IUE dataset and is based on the work of \cite{B&M} who calculated the Mg II index of different stars from IUE high dispersion + large aperture observations. We have reduced the resolution of the HST dataset to match the IUE instrument resolution for application of the same windows. The CUTE dataset is however much lower resolution than the IUE dataset. We have used a wavelength window range of 2760.0 \r{A} - 2780.0 \r{A} for the blue wing and 2828.0 \r{A} - 2848.0 \r{A} for the red wing. For the k and h lines we have used wavelength windows of 2794.0 \r{A} - 2799.0 \r{A} and 2801.0 \r{A} - 2806.0 \r{A} respectively.  During exploration of the impact of CUTE's lower spectral resolution on the Mg II activity index, we found that the data were better described by a $\approx$6.5~\AA\ resolution element (based on the observed profile of the Mg I 2850~\AA\ line), which is lower ($R$~$\approx$~450) than the previously published $R$~$\approx$~750 from CUTE's on-orbit commissioning~\citep{2023Egan}.  Using this lower resolution, we quantified the amount of continuum that 'spills' into the Mg II h and k region from the line wings and the inter-line emission, estimating that the lower resolution biases the Mg II indices observed by CUTE upward by ~30\%.   We corrected for this effect when comparing the activity indices across the three observatories considered in this work.   

\begin{figure*}
    \centering
    \includegraphics[width = \linewidth]{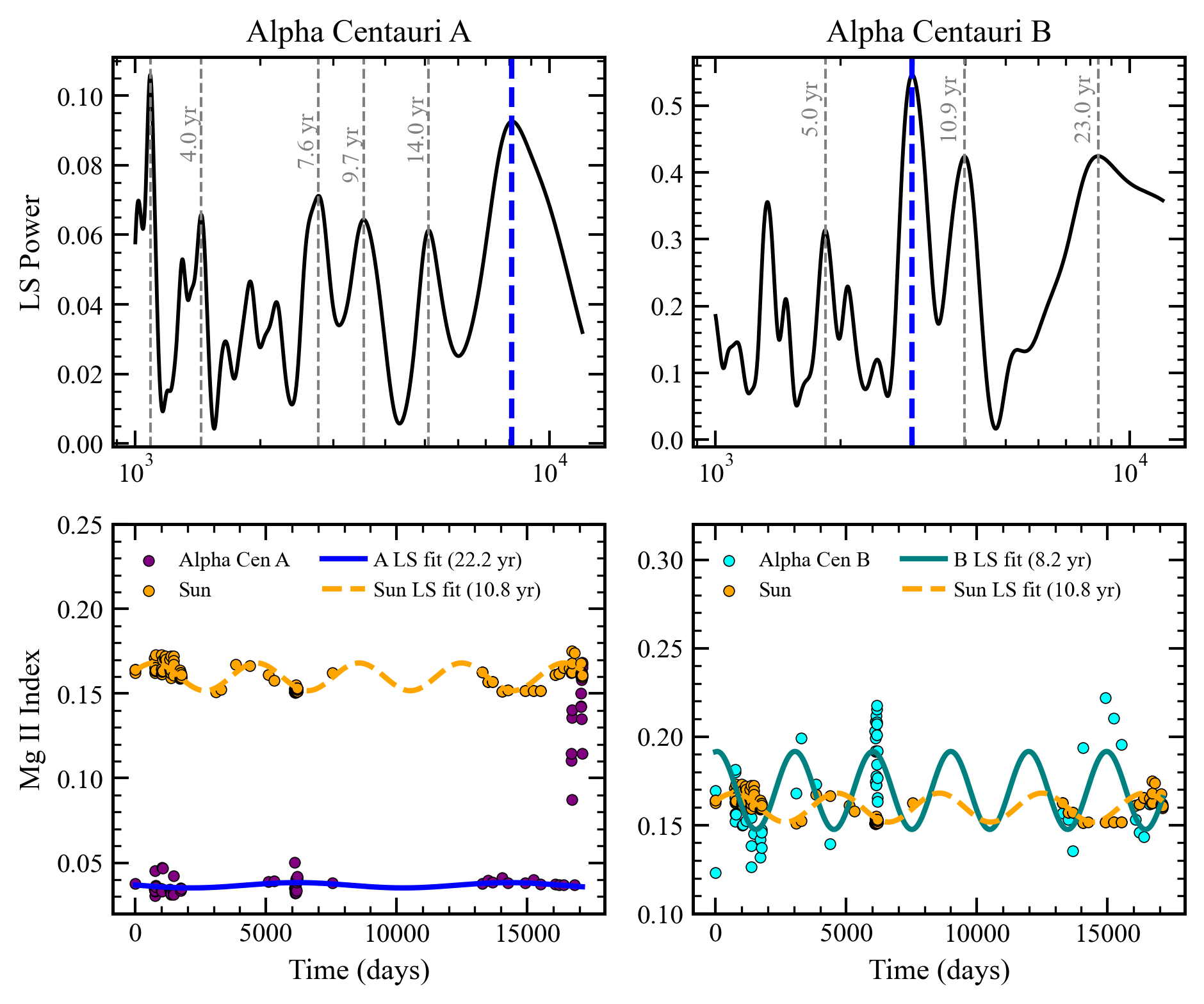}
    \caption{\textbf{Top:} These panels represent the Lomb-Scargle power for different activity cycle power frequencies found in the dataset of Alpha Centauri A and B. The highest power for a period $> 5$ years is marked by the blue dashed line which amounts to 22.2 years for Alpha Centauri A and 8.2 years for Alpha Centauri B. \textbf{Bottom:} These panels represent the sinusoidal fit to the MgII index data for Alpha Centauri A and B. The solar MgII index is overlaid in the figure for reference. The solar MgII index is obtained from the LISIRD website \citep{2019EGUGA..2112479L}.}
    \label{fig:fig_6}
\end{figure*}

\begin{figure*}
    \centering
    \includegraphics[width = \linewidth]{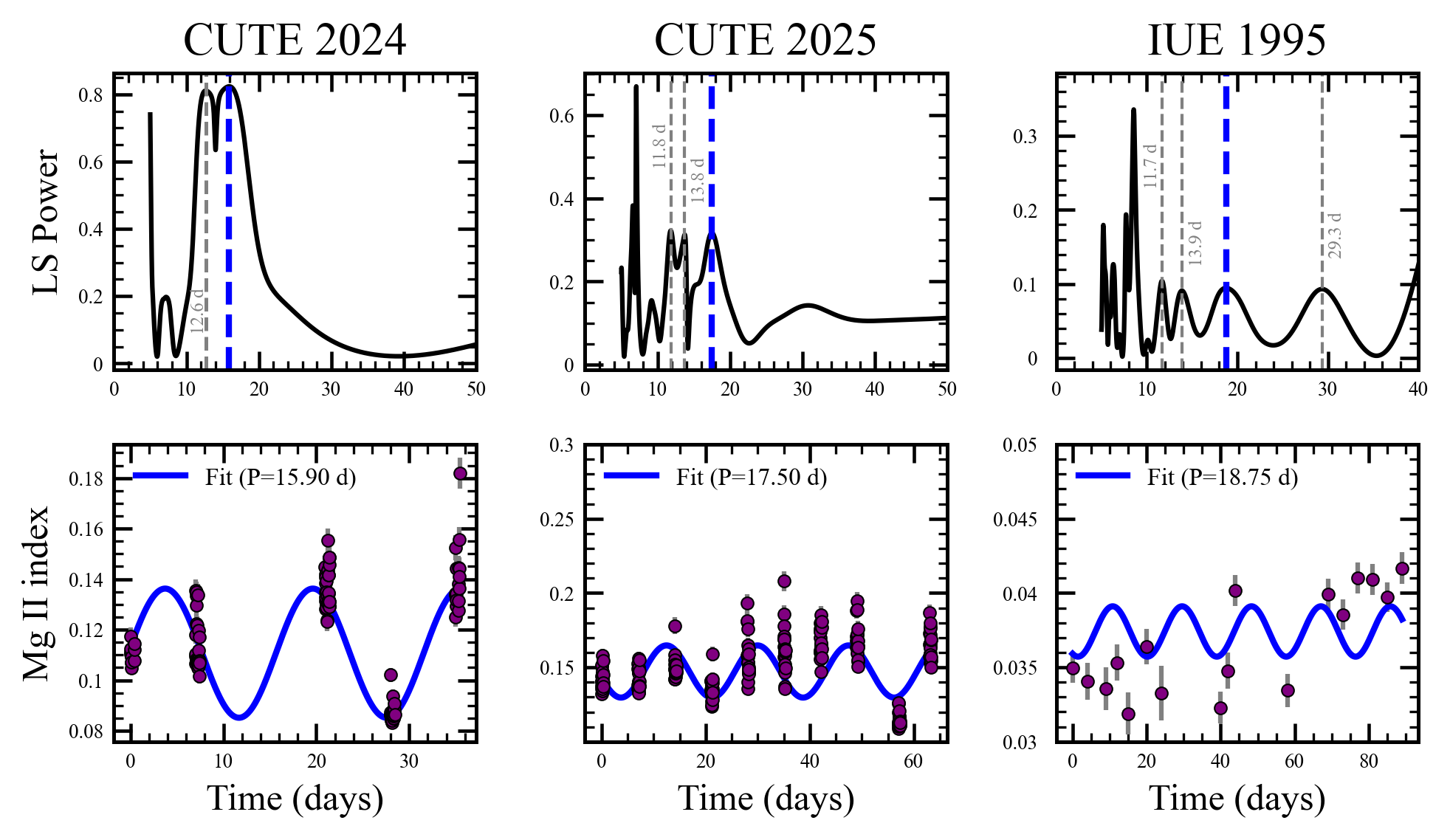}
    \caption{\textbf{Top:} These panels represent Lomb-Scargle power for different rotation period power frequencies found in the CUTE and IUE dataset of Alpha Centauri A. The highest power is marked by the blue dashed line which amounts to 15.9 days for CUTE 2024. Based on the CUTE 2024 analysis, the highest peak between 15 and 20 days was found to be 17.5 days for the CUTE 2025 dataset and 18.75 days for the IUE dataset. \textbf{Bottom:} These panels represent the best period sinusoidal fit to the MgII index data for the CUTE and IUE observations of Alpha Centauri A.} 
    \label{fig:fig_7}
\end{figure*}

We use the Mg II index of Alpha Centauri A and B to investigate both their long-term stellar activity cycles and their short-term rotational modulation. X-ray monitoring has previously been used to establish the stellar cycle periodicities of both stars \citep{Ayres_2023}. Here, we apply the Lomb–Scargle technique to the near-ultraviolet Mg II index to provide an independent chromospheric diagnostic. The Lomb–Scargle periodogram is a spectral analysis method designed to detect periodic signals in unevenly sampled time-series data \citep{1976Lomb,1982Scargle}. It identifies naturally occurring frequencies in a dataset by fitting sinusoidal models at trial frequencies using least-squares minimization, and has been widely used to characterize stellar UV activity cycles from IUE and HST observations \citep{2024Kamgar}. 

By applying the Lomb-Scargle technique to the Alpha Centauri B Mg II time series, we recover a clear stellar cycle with a period of approximately 8 years, in excellent agreement with the cycle derived from X-ray observations \citep{Ayres_2023}. The right column of Figure~\ref{fig:fig_6} shows the corresponding Lomb-Scargle power spectrum. The highest-power peak, marked by the blue dashed line, corresponds to a period of 8.2 $\pm0.75$ years and is fitted to the data in the lower panel. The quoted uncertainty is derived from the half-width at half-maximum (HWHM) of the Lomb-Scargle periodogram peak. 

For comparison, we apply the same analysis to the solar Mg II index obtained from the LASP Interactive Solar IRradience Datacenter (LISIRD) database \citep{2019EGUGA..2112479L}, restricting the data to the same time span and cadence as the Alpha Centauri B observations. This retrieves a solar cycle period of 10.8 years, very close to the canonical 11-year cycle, lending confidence to the reliability of the method and the Alpha Centauri B result.

For Alpha Centauri A, the Lomb-Scargle analysis of the long-term stellar cycle uses only the Mg II indices derived from the IUE and HST datasets, for which the wavelength windows in Equation~\ref{eq:mgii_index} are spectrally resolved and directly comparable. We do not include the CUTE measurements in this long-term analysis because the lower spectral resolution of CUTE does not cleanly separate the Mg II h and k line cores from the surrounding continuum even after the $30\%$ correction. As a result, the CUTE-derived indices include continuum emission in the numerator of Equation~\ref{eq:mgii_index}, producing systematically higher Mg II values (points with Mg II $>0.05$ in Figure~\ref{fig:fig_6}) and introducing a bias that cannot be corrected in a manner consistent with the higher-resolution datasets.

The IUE + HST Lomb-Scargle periodogram for Alpha Centauri A shows its strongest power near a period of 22.2 $\pm6.02$ years (uncertainty is HWHM of Lomb-Scargle periodogram peak), comparable to the $\sim$19-year stellar cycle derived from X-ray monitoring. However, this peak is broad, indicating a low-coherence and poorly constrained periodicity. This behavior is consistent with previous studies showing that Alpha Centauri A exhibits weak chromospheric variability typical of a magnetically quiet G-type star \citep{Ayres_2015,2023Ayres}, and that X-ray diagnostics provide more robust constraints on its stellar cycle \citep{Ayres_2023}.

Although CUTE data cannot be combined with IUE and HST for absolute stellar cycle measurements, they remain well suited for determining the stellar rotation period because the analysis is performed on each CUTE dataset independently. Any resolution-dependent bias is therefore propagated uniformly across each time series and does not affect the relative modulation or the dominant periodicity in the power spectrum.

Previous estimates of the rotation period of Alpha Centauri A from asteroseismology gave a period value of $22.5 \pm5.9$ days \citep{2007Bazot}. Analysis of IUE NUV observations yielded a preliminary value of $23^{+5}_{-2}$ days \citep{1997Rotper}. The CUTE observations provide the highest-cadence NUV monitoring of the star since that time. Using the Mg II index derived consistently from the CUTE spectra, we analyze the chromospheric variability to estimate the stellar rotation period.

For the 2024 CUTE observations, we find a best rotation period of 15.9 $\pm1.67$ days. The 2025 dataset is more strongly affected by spacecraft jitter and pointing instabilities, which often caused the star to drift toward the edge of the slit, leading to intermittent flux loss. These effects likely introduce artificial patterns into the time series. To mitigate low-frequency noise in the 2025 dataset, we restrict the Lomb–Scargle search to periods longer than 15 days partly based on the 2024 CUTE data analysis. With this constraint, the 2025 data yield a primary peak at 17.5 $\pm2.54$ days.

To further validate this result, we analyze the high-cadence IUE 1995 Mg II time series of Alpha Centauri A. Applying period limits of 15–30 days, we recover a strongest peak at 18.75 $\pm2.34$ days. This value is consistent with the CUTE-derived periods. Figure~\ref{fig:fig_7} shows the Lomb–Scargle power spectra (top row) and the corresponding sinusoidal fits (bottom row) for the CUTE and IUE datasets.  Taken together, we find that the CUTE and IUE NUV data sets indicate a stellar rotation period somewhere between 15 - 20 days for Alpha Centauri A. Uncertainties in the rotation periods correspond to the HWHM of the Lomb-Scargle periodogram peak.

The rotation periods derived in this work lie within the uncertainty limit of the period derived from asteroseismology \citep{2007Bazot} but are marginally shorter than the value previously reported by \cite{1997Rotper}, which was based on IUE observations obtained within a $\sim$4-month window from May to August 1995. Because this IUE data were acquired near the end of the mission, they may be affected by increased pointing uncertainties, which can complicate efforts to constrain the true rotation period of Alpha Centauri A. This is reflected in the presence of multiple closely spaced peaks near the maximum-power feature in Figure~\ref{fig:fig_7}, as well as in the poor sinusoidal fit to the time series. 

\section{Discussion and Conclusion}
We have examined the near-ultraviolet (NUV) activity of the G-type star Alpha Centauri A and its K-type companion Alpha Centauri B over a long temporal baseline from 1978 to 2025, using spectral observations from IUE, HST, and CUTE. This unique multi-decade dataset allows us to characterize both short and long term stellar variability in the nearest Sun-like binary system.

The NUV emission of Alpha Centauri A is dominated by low-amplitude variability. Across both photospheric and chromospheric bands, the majority of epochs cluster within $\approx10\%$ of the median flux, with only $\sim20\%$ of measurements deviating beyond this level. Large excursions, with amplitudes exceeding $\approx20\%$ relative to the median, are rare and are associated with chromospheric flaring.

In contrast, Alpha Centauri B exhibits substantially more frequent moderate and strong variability, particularly in the chromosphere. Although approximately $50–60\%$ of the observed epochs remain near the median photospheric level for B, a significant fraction extend beyond $\pm1\sigma$, with multiple excursions exceeding $\pm2\sigma$ and $\pm3\sigma$. These chromospheric enhancements indicate a higher level of magnetic activity for Alpha Centauri B relative to Alpha Centauri A.

We further investigated the rotational modulation of Alpha Centauri A using Lomb–Scargle periodograms of the Mg II index derived from high-cadence CUTE data and archival IUE observations. The 2024 CUTE dataset yields a well-defined modulation near $\sim$16 days, while the 2025 data show additional power at shorter periods likely influenced by spacecraft jitter and sampling effects. Independent analysis of the IUE 1995 time series, restricted to 15–30 days, recovers a primary peak consistent with the CUTE-derived rotation signal. Together, these results support a rotation period of $\approx15–20$ days for Alpha Centauri A, faster than the solar rotation period of 28 days, and demonstrate the utility of near-UV chromospheric diagnostics for probing stellar rotation.

Although Alpha Centauri A is often described as a “solar analogue,” its broader magnetic context indicates that it is not a true analogue for the Sun. Its coronal activity level, quantified by $\log(L_X/L_{\rm bol}) \approx -7.0$ to $-7.2$ is significantly lower than that of B ($\log(L_X/L_{\rm bol}) \approx -6.2$) whose value is similar to the Sun \citep{2014Ayres,2012Robrade}. Likewise, the Ca II h and k chromospheric activity index of Alpha Centauri A is low, with reported $\log R'_{\rm HK} \approx -5.0$ to $-5.1$ \citep{2008Mama,2004Wright}, consistent with an old, magnetically quiet star. Asteroseismic modeling by \cite{2021Salmon} yields a mass of $1.11,M_\odot$ for Alpha Centauri A and $0.94,M_\odot$ for Alpha Centauri B. The higher mass of Alpha Centauri A implies a thinner outer convective envelope, which is expected to reduce magnetic braking efficiency relative to the Sun. Consequently, a faster rotation period for Alpha Centauri A than for the present-day Sun is not unexpected and does not indicate enhanced magnetic activity. Instead, the combination of higher mass, and reduced convective depth likely explains why Alpha Centauri A is both rotationally distinct from the Sun and magnetically quieter, despite its superficial similarity in spectral type.

From an exoplanet perspective, the measured stellar variability envelopes set practical detection thresholds. For an Earth-sized rocky planet orbiting Alpha Centauri A, intrinsic planetary NUV variability would need to exceed the stellar envelope ($\gtrsim20\%$ peak-to-median amplitude) and/or exhibit temporal behavior uncorrelated with stellar activity to be confidently distinguished from stellar-driven variability. For a comparable planet orbiting Alpha Centauri B, the required threshold is larger, $\gtrsim30–40\%$ relative to the median, reflecting the star’s stronger chromospheric variability.  

In reflected-light spectroscopy, the observed planetary signal is proportional to the product of the stellar spectral irradiance and the planet’s wavelength-dependent reflectance. As a result, fractional variability in the stellar near-UV continuum propagates directly into comparable fractional uncertainty in the inferred depth of ozone absorption bands unless the stellar spectrum is known independently. The $\sim20\%$ peak-to-median NUV variability envelope measured for Alpha Centauri A therefore corresponds to a systematic uncertainty of similar magnitude in retrieved O$_3$ band depths, and hence in inferred O$_2$ abundance, under continuum-limited conditions. In the absence of simultaneous NUV stellar monitoring, this sets a practical floor on ozone retrieval precision for Earth-like planets orbiting nearby solar-type stars.

In conclusion, the results presented in this paper provide empirically grounded bounds on stellar-driven NUV variability for the nearest Sun-like and K-type stars in the Alpha Centauri system. These constraints establish a critical reference framework for interpreting reflected-light observations of terrestrial exoplanet atmospheres and for assessing the detectability of ozone and other biosignature-related features with future facilities such as the Habitable Worlds Observatory.

\section{ACKNOWLEDGMENTS}       
This work was supported by NASA/APRA awards NNX17AI8G, 80NSSC21K1667, and 80NSSC25K7312 (PI~--~K. France) to LASP at the University of Colorado, Boulder. DJW was supported by NASA ADSPS  grant award number 80NSSC23K1474. 

This paper employs a list of IUE and HST datasets hosted on the MAST archive contained in the collections:\dataset[IUE-MAST]{https://doi.org/10.17909/p38z-4y59} and \dataset[HST-MAST]{https://doi.org/10.17909/dk8s-n214}.

 
\FloatBarrier
\bibliography{bibtex}{}

@INPROCEEDINGS{2021Nico,
       author = {{Nell}, Nicholas and {DeCicco}, Nicholas and {Ulrich}, Stefan and {France}, Kevin and {Fleming}, Brian},
        title = "{Development and characterization of the CCD detector for the Colorado Ultraviolet Transit Experiment (CUTE) cubesat}",
    booktitle = {UV, X-Ray, and Gamma-Ray Space Instrumentation for Astronomy XXII},
         year = 2021,
       editor = {{Siegmund}, Oswald H.},
       series = {Society of Photo-Optical Instrumentation Engineers (SPIE) Conference Series},
       volume = {11821},
        month = aug,
          eid = {1182117},
        pages = {1182117},
          doi = {10.1117/12.2594743},
       adsurl = {https://ui.adsabs.harvard.edu/abs/2021SPIE11821E..17N}
}

@ARTICLE{2023Egan,
       author = {{Egan}, Arika and {Nell}, Nicholas and {Suresh}, Ambily and {France}, Kevin and {Fleming}, Brian and {Sreejith}, Aickara Gopinathan and {Lambert}, Julian and {DeCicco}, Nicholas},
        title = "{The On-orbit Performance of the Colorado Ultraviolet Transit Experiment Mission}",
      journal = {\aj},
         year = 2023,
        month = feb,
       volume = {165},
       number = {2},
          eid = {64},
        pages = {64},
          doi = {10.3847/1538-3881/aca8a3},
archivePrefix = {arXiv},
       eprint = {2301.01307},
 primaryClass = {astro-ph.IM},
       adsurl = {https://ui.adsabs.harvard.edu/abs/2023AJ....165...64E}
}

@ARTICLE{2023Ayres,
       author = {{Ayres}, Thomas},
        title = "{The Cycles of Alpha Centauri: Double Dipping of AB}",
      journal = {\aj},
         year = 2023,
        month = nov,
       volume = {166},
       number = {5},
          eid = {212},
        pages = {212},
          doi = {10.3847/1538-3881/acfef5},
       adsurl = {https://ui.adsabs.harvard.edu/abs/2023AJ....166..212A}
}

@BOOK{2020Astro,
       author = {Astro2020},
        title = "{Pathways to Discovery in Astronomy and Astrophysics for the 2020s}",
         year = {2021},
          doi = {10.17226/26141},
       adsurl = {https://ui.adsabs.harvard.edu/abs/2021pdaa.book.....N}
}

@ARTICLE{tuchow25,
       author = {{Tuchow}, Noah W. and {Harada}, Caleb K. and {Mamajek}, Eric E. and {Tanner}, Angelle and {Hinkel}, Natalie R. and {Belikov}, Ruslan and {Sirbu}, Dan and {Ciardi}, David R. and {Stark}, Christopher C. and {Morgan}, Rhonda M. and {Savransky}, Dmitri and {Turmon}, Michael},
        title = "{HWO Target Stars and Systems: A Prioritized Community List of Potential Stellar Targets for the Habitable Worlds Observatory's ExoEarth Survey}",
      journal = {\pasp},
         year = 2025,
        month = oct,
       volume = {137},
       number = {10},
          eid = {104402},
        pages = {104402},
          doi = {10.1088/1538-3873/ae0a81},
       adsurl = {https://ui.adsabs.harvard.edu/abs/2025PASP..137j4402T}
}

@ARTICLE{peacock25,
       author = {{Peacock}, Sarah and {Wilson}, David J. and {Richey-Yowell}, Tyler and {Tuchow}, Noah W. and {France}, Kevin and {Caballero}, Jos{\'e} A. and {Spinelli}, Riccardo and {Corrales}, L{\'\i}a and {Zelakiewicz}, Aiden S. and {Redfield}, Seth and {Rockcliffe}, Keighley and {Youngblood}, Allison and {Froning}, Cynthia S. and {Duvvuri}, Girish M. and {Binder}, Breanna A. and {Hinkel}, Natalie R. and {Mamajek}, Eric E.},
        title = "{HWO Target Stars and Systems: A Survey of Archival UV and X-ray Data}",
      journal = {arXiv e-prints},
         year = 2025,
        month = sep,
          eid = {arXiv:2509.08999},
        pages = {arXiv:2509.08999},
          doi = {10.48550/arXiv.2509.08999},
archivePrefix = {arXiv},
       eprint = {2509.08999},
 primaryClass = {astro-ph.SR},
       adsurl = {https://ui.adsabs.harvard.edu/abs/2025arXiv250908999P}
}

@ARTICLE{totton25,
       author = {{Krissansen-Totton}, Joshua and {Ulses}, Anna Grace and {Frissell}, Maxwell and {Gilbert-Janizek}, Samantha and {Young}, Amber and {Lustig-Yaeger}, Jacob and {Robinson}, Tyler and {Olson}, Stephanie and {Alei}, Eleonora and {Arney}, Giada and {Hagee}, Celeste and {Harman}, Chester and {Hinkel}, Natalie and {Lafleche}, Emilie and {Latouf}, Natasha and {Mandell}, Avi and {Moussa}, Mark M. and {Parenteau}, Niki and {Ranjan}, Sukrit and {Russell}, Blair and {Schwieterman}, Edward W. and {Sousa-Silva}, Clara and {Tokadjian}, Armen and {Wogan}, Nicholas},
        title = "{Wavelength Requirements for Life Detection via Reflected Light Spectroscopy of Rocky Exoplanets}",
      journal = {arXiv e-prints},
         year = 2025,
        month = jul,
          eid = {arXiv:2507.14771},
        pages = {arXiv:2507.14771},
          doi = {10.48550/arXiv.2507.14771},
archivePrefix = {arXiv},
       eprint = {2507.14771},
 primaryClass = {astro-ph.EP},
       adsurl = {https://ui.adsabs.harvard.edu/abs/2025arXiv250714771K}
}

@inproceedings{cute2025db,
author = {Dolon Bhattacharyya and Kevin France and Sierra Flynn and Sebastian Escobar and Arika Egan and Max Fairchild},
title = {{The CUTE CubeSat mission: insights into instrument and spacecraft performance evolution over 4(!) years}},
volume = {13625},
booktitle = {UV, X-Ray, and Gamma-Ray Space Instrumentation for Astronomy XXIV},
editor = {Oswald H. Siegmund and Keri Hoadley},
organization = {International Society for Optics and Photonics},
publisher = {SPIE},
pages = {136251M},
year = {2025},
doi = {10.1117/12.3064142},
URL = {https://doi.org/10.1117/12.3064142}
}

@article{Ayres_2009,
doi = {10.1088/0004-637X/696/2/1931},
url = {https://doi.org/10.1088/0004-637X/696/2/1931},
year = {2009},
month = {apr},
publisher = {The American Astronomical Society},
volume = {696},
number = {2},
pages = {1931},
author = {Ayres, Thomas R.},
title = {THE CYCLES OF α CENTAURI},
journal = {The Astrophysical Journal}
}

@article{Ayres_2023,
doi = {10.3847/1538-3881/acfef5},
url = {https://doi.org/10.3847/1538-3881/acfef5},
year = {2023},
month = {nov},
publisher = {The American Astronomical Society},
volume = {166},
number = {5},
pages = {212},
author = {Ayres, Thomas},
title = {The Cycles of Alpha Centauri: Double Dipping of AB},
journal = {The Astronomical Journal}
}

@article{Ayres_2015,
doi = {10.1088/0004-6256/149/2/58},
url = {https://doi.org/10.1088/0004-6256/149/2/58},
year = {2015},
month = {jan},
publisher = {The American Astronomical Society},
volume = {149},
number = {2},
pages = {58},
author = {Ayres, Thomas R.},
title = {THE FAR-ULTRAVIOLET UPS AND DOWNS OF ALPHA CENTAURI},
journal = {The Astronomical Journal}
}

@ARTICLE{2023France,
       author = {{France}, Kevin and {Fleming}, Brian and {Egan}, Arika and {Desert}, Jean-Michel and {Fossati}, Luca and {Koskinen}, Tommi T. and {Nell}, Nicholas and {Petit}, Pascal and {Vidotto}, Aline A. and {Beasley}, Matthew and {DeCicco}, Nicholas and {Sreejith}, Aickara Gopinathan and {Suresh}, Ambily and {Baumert}, Jared and {Cauley}, P. Wilson and {Villarreal D'Angelo}, Carolina and {Hoadley}, Keri and {Kane}, Robert and {Kohnert}, Richard and {Lambert}, Julian and {Ulrich}, Stefan},
        title = "{The Colorado Ultraviolet Transit Experiment Mission Overview}",
      journal = {\aj},
         year = 2023,
        month = feb,
       volume = {165},
       number = {2},
          eid = {63},
        pages = {63},
          doi = {10.3847/1538-3881/aca8a2},
archivePrefix = {arXiv},
       eprint = {2301.02250},
 primaryClass = {astro-ph.IM},
       adsurl = {https://ui.adsabs.harvard.edu/abs/2023AJ....165...63F}
}

@ARTICLE{B&M,
       author = {{Buccino}, A.~P. and {Mauas}, P.~J.~D.},
        title = "{Mg II h+k emission lines as stellar activity indicators of main sequence F-K stars}",
      journal = {\aap},
         year = 2008,
        month = jun,
       volume = {483},
       number = {3},
        pages = {903-910},
          doi = {10.1051/0004-6361:20078925},
archivePrefix = {arXiv},
       eprint = {0804.1101},
 primaryClass = {astro-ph},
       adsurl = {https://ui.adsabs.harvard.edu/abs/2008A&A...483..903B}
}

@ARTICLE{HandS,
       author = {{Heath}, D.~F. and {Schlesinger}, B.~M.},
        title = "{The Mg 280-nm doublet as a monitor of changes in solar ultraviolet irradiance}",
      journal = {\jgr},
         year = 1986,
        month = jul,
       volume = {91},
       number = {D8},
        pages = {8672-8682},
          doi = {10.1029/JD091iD08p08672},
       adsurl = {https://ui.adsabs.harvard.edu/abs/1986JGR....91.8672H}
}

@ARTICLE{1976Lomb,
       author = {{Lomb}, N.~R.},
        title = "{Least-Squares Frequency Analysis of Unequally Spaced Data}",
      journal = {\apss},
         year = 1976,
        month = feb,
       volume = {39},
       number = {2},
        pages = {447-462},
          doi = {10.1007/BF00648343},
       adsurl = {https://ui.adsabs.harvard.edu/abs/1976Ap&SS..39..447L}
}

@ARTICLE{1982Scargle,
       author = {{Scargle}, J.~D.},
        title = "{Studies in astronomical time series analysis. II. Statistical aspects of spectral analysis of unevenly spaced data.}",
      journal = {\apj},
         year = 1982,
        month = dec,
       volume = {263},
        pages = {835-853},
          doi = {10.1086/160554},
       adsurl = {https://ui.adsabs.harvard.edu/abs/1982ApJ...263..835S}
}

@ARTICLE{2024Kamgar,
       author = {{Kamgar}, Leo and {France}, Kevin and {Youngblood}, Allison},
        title = "{Time Variability of FUV Emission from Cool Stars on Multi-year Timescales}",
      journal = {\pasp},
         year = 2024,
        month = feb,
       volume = {136},
       number = {2},
          eid = {024202},
        pages = {024202},
          doi = {10.1088/1538-3873/ad119f},
       adsurl = {https://ui.adsabs.harvard.edu/abs/2024PASP..136b4202K}
}

@INPROCEEDINGS{2019EGUGA..2112479L,
       author = {{Leise}, Hunter and {Baltzer}, Tom and {Wilson}, Anne and {Lindholm}, Doug and {Snow}, Martin and {Woodraska}, Don and {B{\'e}land}, St{\'e}phane and {Coddington}, Odele and {Pankratz}, Chris},
        title = "{LASP Interactive Solar IRradiance Datacenter (LISIRD)}",
    booktitle = {EGU General Assembly Conference Abstracts},
         year = 2019,
       series = {EGU General Assembly Conference Abstracts},
        month = apr,
          eid = {12479},
        pages = {12479},
       adsurl = {https://ui.adsabs.harvard.edu/abs/2019EGUGA..2112479L}
}

@INPROCEEDINGS{1997Rotper,
       author = {{Jay}, J.~E. and {Guinan}, E.~F. and {Morgan}, N.~D. and {Messina}, S. and {Jassour}, D.},
        title = "{Rotation and Stellar Activity of the Stars of the {\ensuremath{\alpha}} Centauri Triple Star System}",
    booktitle = {American Astronomical Society Meeting Abstracts \#189},
         year = 1997,
       series = {American Astronomical Society Meeting Abstracts},
       volume = {189},
        month = jan,
          eid = {120.04},
        pages = {120.04},
       adsurl = {https://ui.adsabs.harvard.edu/abs/1997AAS...18912004J}
}

@ARTICLE{bryson20,
       author = {{Bryson}, S. and {Coughlin}, J. and {Batalha}, N.~M. and {Berger}, T. and {Huber}, D. and {Burke}, C. and {Dotson}, J. and {Mullally}, S.~E.},
        title = "{A Probabilistic Approach to Kepler Completeness and Reliability for Exoplanet Occurrence Rates}",
      journal = {\aj},
         year = 2020,
        month = jun,
       volume = {159},
       number = {6},
          eid = {279},
        pages = {279},
          doi = {10.3847/1538-3881/ab8a30},
archivePrefix = {arXiv},
       eprint = {1906.03575},
 primaryClass = {astro-ph.EP},
       adsurl = {https://ui.adsabs.harvard.edu/abs/2020AJ....159..279B}
}

@ARTICLE{Krissansen-Totton18,
       author = {{Krissansen-Totton}, Joshua and {Garland}, Ryan and {Irwin}, Patrick and {Catling}, David C.},
        title = "{Detectability of Biosignatures in Anoxic Atmospheres with the James Webb Space Telescope: A TRAPPIST-1e Case Study}",
      journal = {\aj},
         year = 2018,
        month = sep,
       volume = {156},
       number = {3},
          eid = {114},
        pages = {114},
          doi = {10.3847/1538-3881/aad564},
archivePrefix = {arXiv},
       eprint = {1808.08377},
 primaryClass = {astro-ph.EP},
       adsurl = {https://ui.adsabs.harvard.edu/abs/2018AJ....156..114K}
}

@ARTICLE{arney16,
       author = {{Arney}, Giada and {Domagal-Goldman}, Shawn D. and {Meadows}, Victoria S. and {Wolf}, Eric T. and {Schwieterman}, Edward and {Charnay}, Benjamin and {Claire}, Mark and {H{\'e}brard}, Eric and {Trainer}, Melissa G.},
        title = "{The Pale Orange Dot: The Spectrum and Habitability of Hazy Archean Earth}",
      journal = {Astrobiology},
         year = 2016,
        month = nov,
       volume = {16},
       number = {11},
        pages = {873-899},
          doi = {10.1089/ast.2015.1422},
archivePrefix = {arXiv},
       eprint = {1610.04515},
 primaryClass = {astro-ph.EP},
       adsurl = {https://ui.adsabs.harvard.edu/abs/2016AsBio..16..873A}
}

@ARTICLE{damiano2023,
       author = {{Damiano}, Mario and {Hu}, Renyu and {Mennesson}, Bertrand},
        title = "{Reflected Spectroscopy of Small Exoplanets. III. Probing the UV Band to Measure Biosignature Gases}",
      journal = {\aj},
         year = 2023,
        month = oct,
       volume = {166},
       number = {4},
          eid = {157},
        pages = {157},
          doi = {10.3847/1538-3881/acefd3},
archivePrefix = {arXiv},
       eprint = {2308.08490},
 primaryClass = {astro-ph.EP},
       adsurl = {https://ui.adsabs.harvard.edu/abs/2023AJ....166..157D}
}

@ARTICLE{loyd16,
       author = {{Loyd}, R.~O.~P. and {France}, Kevin and {Youngblood}, Allison and {Schneider}, Christian and {Brown}, Alexander and {Hu}, Renyu and {Linsky}, Jeffrey and {Froning}, Cynthia S. and {Redfield}, Seth and {Rugheimer}, Sarah and {Tian}, Feng},
        title = "{The MUSCLES Treasury Survey. III. X-Ray to Infrared Spectra of 11 M and K Stars Hosting Planets}",
      journal = {\apj},
         year = 2016,
        month = jun,
       volume = {824},
       number = {2},
          eid = {102},
        pages = {102},
          doi = {10.3847/0004-637X/824/2/102},
archivePrefix = {arXiv},
       eprint = {1604.04776},
 primaryClass = {astro-ph.SR},
       adsurl = {https://ui.adsabs.harvard.edu/abs/2016ApJ...824..102L}
}

@manual{STIS_IHB,
  author       = {{Space Telescope Science Institute}},
  title        = {STIS Instrument Handbook, Version 20.0},
  year         = {2023},
  organization = {STScI},
  address      = {Baltimore, MD},
  url          = {https://hst-docs.stsci.edu/stisihb}
}

@ARTICLE{2009K_flares,
       author = {{Kowalski}, Adam F. and {Hawley}, Suzanne L. and {Hilton}, Eric J. and {Becker}, Andrew C. and {West}, Andrew A. and {Bochanski}, John J. and {Sesar}, Branimir},
        title = "{M Dwarfs in Sloan Digital Sky Survey Stripe 82: Photometric Light Curves and Flare Rate Analysis}",
      journal = {\aj},
         year = 2009,
        month = aug,
       volume = {138},
       number = {2},
        pages = {633-648},
          doi = {10.1088/0004-6256/138/2/633},
archivePrefix = {arXiv},
       eprint = {0906.2030},
 primaryClass = {astro-ph.SR},
       adsurl = {https://ui.adsabs.harvard.edu/abs/2009AJ....138..633K}
}

@ARTICLE{2010K_flares,
       author = {{Kowalski}, Adam F. and {Hawley}, Suzanne L. and {Holtzman}, Jon A. and {Wisniewski}, John P. and {Hilton}, Eric J.},
        title = "{A White Light Megaflare on the dM4.5e Star YZ CMi}",
      journal = {\apjl},
         year = 2010,
        month = may,
       volume = {714},
       number = {1},
        pages = {L98-L102},
          doi = {10.1088/2041-8205/714/1/L98},
archivePrefix = {arXiv},
       eprint = {1003.3057},
 primaryClass = {astro-ph.SR},
       adsurl = {https://ui.adsabs.harvard.edu/abs/2010ApJ...714L..98K}
}

@ARTICLE{2019K_flares,
       author = {{Kowalski}, Adam F. and {Wisniewski}, John P. and {Hawley}, Suzanne L. and {Osten}, Rachel A. and {Brown}, Alexander and {Fari{\~n}a}, Cecilia and {Valenti}, Jeff A. and {Brown}, Stephen and {Xilouris}, Manolis and {Schmidt}, Sarah J. and {Johns-Krull}, Christopher},
        title = "{The Near-ultraviolet Continuum Radiation in the Impulsive Phase of HF/GF-type dMe Flares. I. Data}",
      journal = {\apj},
         year = 2019,
        month = feb,
       volume = {871},
       number = {2},
          eid = {167},
        pages = {167},
          doi = {10.3847/1538-4357/aaf058},
archivePrefix = {arXiv},
       eprint = {1811.04021},
 primaryClass = {astro-ph.SR},
       adsurl = {https://ui.adsabs.harvard.edu/abs/2019ApJ...871..167K}
}

@article{Kleint2016ApJ81688,
  author  = {Kleint, L. and Heinzel, P. and Judge, P. and Krucker, S.},
  title   = {Continuum Enhancements in the Ultraviolet, the Visible, and the Infrared during the X1 Flare on 2014 March 29},
  journal = {ApJ},
  volume  = {816},
  pages   = {88},
  year    = {2016}
}

@article{Kerr2015AA582A50,
  author  = {Kerr, G. S. and Sim{\~o}es, P. J. A. and Qiu, J. and Fletcher, L.},
  title   = {{IRIS} Observations of the {Mg II} h and k Lines during a Solar Flare},
  journal = {A\&A},
  volume  = {582},
  pages   = {A50},
  year    = {2015}
}

@ARTICLE{1999Viereck,
       author = {{Viereck}, Rodney A. and {Puga}, Lawrence C.},
        title = "{The NOAA Mg II core-to-wing solar index: Construction of a 20-year time series of chromospheric variability from multiple satellites}",
      journal = {\jgr},
         year = 1999,
        month = may,
       volume = {104},
       number = {A5},
        pages = {9995-10006},
          doi = {10.1029/1998JA900163},
       adsurl = {https://ui.adsabs.harvard.edu/abs/1999JGR...104.9995V}
}

@ARTICLE{2019ESnow,
       author = {{Snow}, M. and {Machol}, J. and {Viereck}, R. and {Woods}, T. and {Weber}, M. and {Woodraska}, D. and {Elliott}, J.},
        title = "{A Revised Magnesium II Core-to-Wing Ratio From SORCE SOLSTICE}",
      journal = {Earth and Space Science},
         year = 2019,
        month = nov,
       volume = {6},
       number = {11},
        pages = {2106-2114},
          doi = {10.1029/2019EA000652},
       adsurl = {https://ui.adsabs.harvard.edu/abs/2019E&SS....6.2106S}
}

@article{Donnelly1988_AdvSpaceRes,
  author    = {Donnelly, R. F.},
  title     = {The solar UV core-to-wing ratio from the {NOAA} satellite during the rise of solar cycle 22},
  journal   = {Advances in Space Research},
  volume    = {8},
  pages     = {777--780},
  year      = {1988},
  doi       = {10.1016/0273-1177(88)90174-3}
}

@inproceedings{Skupin2005_ESASP572,
  author       = {Skupin, J. and Weber, M. and Bovensmann, H. and Burrows, J. P.},
  title        = {The Mg II Solar Activity Proxy Indicator Derived from GOME and SCIAMACHY},
  booktitle    = {Proceedings of the ENVISAT I\& ERS Symposium, {ESA} Special Publication SP-572},
  year         = {2005},
  pages        = {179--182},
  organization = {European Space Agency},
  address      = {Noordwijk, The Netherlands}
}

@ARTICLE{2014Ayres,
       author = {{Ayres}, Thomas R.},
        title = "{The Ups and Downs of {\ensuremath{\alpha}} Centauri}",
      journal = {\aj},
         year = 2014,
        month = mar,
       volume = {147},
       number = {3},
          eid = {59},
        pages = {59},
          doi = {10.1088/0004-6256/147/3/59},
archivePrefix = {arXiv},
       eprint = {1401.0847},
 primaryClass = {astro-ph.SR},
       adsurl = {https://ui.adsabs.harvard.edu/abs/2014AJ....147...59A}
}

@ARTICLE{2012Robrade,
       author = {{Robrade}, J. and {Schmitt}, J.~H.~M.~M. and {Favata}, F.},
        title = "{Coronal activity cycles in nearby G and K stars. XMM-Newton monitoring of 61 Cygni and {\ensuremath{\alpha}} Centauri}",
      journal = {\aap},
         year = 2012,
        month = jul,
       volume = {543},
          eid = {A84},
        pages = {A84},
          doi = {10.1051/0004-6361/201219046},
archivePrefix = {arXiv},
       eprint = {1205.3627},
 primaryClass = {astro-ph.SR},
       adsurl = {https://ui.adsabs.harvard.edu/abs/2012A&A...543A..84R}
}

@ARTICLE{2008Mama,
       author = {{Mamajek}, Eric E. and {Hillenbrand}, Lynne A.},
        title = "{Improved Age Estimation for Solar-Type Dwarfs Using Activity-Rotation Diagnostics}",
      journal = {\apj},
         year = 2008,
        month = nov,
       volume = {687},
       number = {2},
        pages = {1264-1293},
          doi = {10.1086/591785},
archivePrefix = {arXiv},
       eprint = {0807.1686},
 primaryClass = {astro-ph},
       adsurl = {https://ui.adsabs.harvard.edu/abs/2008ApJ...687.1264M}
}

@ARTICLE{2004Wright,
       author = {{Wright}, J.~T. and {Marcy}, G.~W. and {Butler}, R. Paul and {Vogt}, S.~S.},
        title = "{Chromospheric Ca II Emission in Nearby F, G, K, and M Stars}",
      journal = {\apjs},
         year = 2004,
        month = jun,
       volume = {152},
       number = {2},
        pages = {261-295},
          doi = {10.1086/386283},
archivePrefix = {arXiv},
       eprint = {astro-ph/0402582},
 primaryClass = {astro-ph},
       adsurl = {https://ui.adsabs.harvard.edu/abs/2004ApJS..152..261W}
}

@ARTICLE{2021Salmon,
       author = {{Salmon}, S.~J.~A.~J. and {Van Grootel}, V. and {Buldgen}, G. and {Dupret}, M.-A. and {Eggenberger}, P.},
        title = "{Reinvestigating {\ensuremath{\alpha}} Centauri AB in light of asteroseismic forward and inverse methods}",
      journal = {\aap},
         year = 2021,
        month = feb,
       volume = {646},
          eid = {A7},
        pages = {A7},
          doi = {10.1051/0004-6361/201937174},
archivePrefix = {arXiv},
       eprint = {2011.14932},
 primaryClass = {astro-ph.SR},
       adsurl = {https://ui.adsabs.harvard.edu/abs/2021A&A...646A...7S}
}

@ARTICLE{2024Arika,
       author = {{Egan}, Arika and {France}, Kevin and {Sreejith}, Aickara Gopinathan and {Fossati}, Luca and {Koskinen}, Tommi and {Fleming}, Brian and {Nell}, Nicholas and {Suresh}, Ambily and {Cauley}, P. Wilson and {Desert}, Jean-Michel and {Petit}, Pascal and {Vidotto}, Aline A.},
        title = "{Colorado Ultraviolet Transit Experiment Near-ultraviolet Transmission Spectroscopy of the Ultrahot Jupiter KELT-9b}",
      journal = {\aj},
         year = 2024,
        month = sep,
       volume = {168},
       number = {3},
          eid = {108},
        pages = {108},
          doi = {10.3847/1538-3881/ad61e5},
archivePrefix = {arXiv},
       eprint = {2407.13656},
 primaryClass = {astro-ph.EP},
       adsurl = {https://ui.adsabs.harvard.edu/abs/2024AJ....168..108E}
}

@ARTICLE{2023Sreejith,
       author = {{Sreejith}, A.~G. and {France}, Kevin and {Fossati}, Luca and {Koskinen}, Tommi T. and {Egan}, Arika and {Cauley}, P. Wilson and {Cubillos}, Patricio. E. and {Ambily}, S. and {Huang}, Chenliang and {Lavvas}, Panayotis and {Fleming}, Brian T. and {Desert}, Jean-Michel and {Nell}, Nicholas and {Petit}, Pascal and {Vidotto}, Aline},
        title = "{CUTE Reveals Escaping Metals in the Upper Atmosphere of the Ultrahot Jupiter WASP-189b}",
      journal = {\apjl},
         year = 2023,
        month = sep,
       volume = {954},
       number = {1},
          eid = {L23},
        pages = {L23},
          doi = {10.3847/2041-8213/acef1c},
archivePrefix = {arXiv},
       eprint = {2308.05726},
 primaryClass = {astro-ph.EP},
       adsurl = {https://ui.adsabs.harvard.edu/abs/2023ApJ...954L..23S}
}

@ARTICLE{2022Sreejith,
       author = {{Sreejith}, A.~G. and {Fossati}, Luca and {Ambily}, S. and {Egan}, Arika and {Nell}, Nicholas and {France}, Kevin and {Fleming}, Brian T. and {Haas}, Stephanie and {Chambliss}, Michael and {DeCicco}, Nicholas and {Steller}, Manfred},
        title = "{The Autonomous Data Reduction Pipeline for the Cute Mission}",
      journal = {\pasp},
         year = 2022,
        month = nov,
       volume = {134},
       number = {1041},
          eid = {114506},
        pages = {114506},
          doi = {10.1088/1538-3873/aca17d},
archivePrefix = {arXiv},
       eprint = {2211.03875},
 primaryClass = {astro-ph.IM},
       adsurl = {https://ui.adsabs.harvard.edu/abs/2022PASP..134k4506S}
}

@dataset{Bhattacharyya2026AlphaCenNUV,
  author       = {Bhattacharyya, Dolon and France, Kevin and Wilson, David},
  title        = {{NUV Wavelength Integrated Flux of Alpha Centauri A and B from IUE, HST, and CUTE}},
  year         = {2026},
  publisher    = {Zenodo},
  doi          = {10.5281/zenodo.18498525},
  url          = {https://doi.org/10.5281/zenodo.18498525}
}

@ARTICLE{2007Bazot,
       author = {{Bazot}, M. and {Bouchy}, F. and {Kjeldsen}, H. and {Charpinet}, S. and {Laymand}, M. and {Vauclair}, S.},
        title = "{Asteroseismology of {\ensuremath{\alpha}} Centauri A. Evidence of rotational splitting}",
      journal = {\aap},
         year = 2007,
        month = jul,
       volume = {470},
       number = {1},
        pages = {295-302},
          doi = {10.1051/0004-6361:20065694},
archivePrefix = {arXiv},
       eprint = {0706.1682},
 primaryClass = {astro-ph},
       adsurl = {https://ui.adsabs.harvard.edu/abs/2007A&A...470..295B}
}

@misc{nasa_hwo,
  author       = {{NASA}},
  title        = {Habitable Worlds Observatory},
  year         = 2026,
  url          = {https://science.nasa.gov/astrophysics/programs/habitable-worlds-observatory/},
  note         = {Accessed: 2026-04-20}
}

@ARTICLE{2026Feinberg,
       author = {{Feinberg}, Lee D. and {Sitarski}, Breann N. and {McElwain}, Michael W. and {Arney}, Giada and {Baker}, Caleb and {Bolcar}, Matthew R. and {Levine}, Marie and {Liu}, Alice and {Mennesson}, Bertrand and {Roberge}, Aki and {Smith}, J. Scott and {Zhao}, Feng and {Ziemer}, John},
        title = "{Habitable Worlds Observatory's Concept and Technology Maturation: Initial Feasibility and Trade Space Exploration}",
      journal = {arXiv e-prints},
         year = 2026,
        month = jan,
          eid = {arXiv:2601.11803},
        pages = {arXiv:2601.11803},
          doi = {10.48550/arXiv.2601.11803},
archivePrefix = {arXiv},
       eprint = {2601.11803},
 primaryClass = {astro-ph.IM},
       adsurl = {https://ui.adsabs.harvard.edu/abs/2026arXiv260111803F}
}
\bibliographystyle{aasjournal}
\end{document}